\documentclass[a4paper]{amsart}

\usepackage{a4wide}
\usepackage[american]{babel}
\usepackage{graphicx}
	\graphicspath{{./}{figures/}}
\usepackage[colorlinks=true,linkcolor=blue,citecolor=red]{hyperref}
\usepackage{xcolor}
\usepackage{mathtools,bbm}

\newtheorem{remark}{Remark}[section]

\DeclarePairedDelimiter{\abs}{\lvert}{\rvert}
\DeclarePairedDelimiter{\avefences}{\langle}{\rangle}

\newcommand{\card}{\operatorname{card}}

\newcommand{\figscale}{0.8}
\newcommand{\norm}[1]{\lVert#1\rVert_\infty}
\newcommand{\R}{\mathbb{R}}
\newcommand{\sgn}{\operatorname{sgn}}
\newcommand{\supp}{\operatorname{supp}}

\title[Multiagent systems with symmetry breaking]{Reducing complexity of multiagent systems with symmetry breaking: an application to opinion dynamics with polls}

\author{	Emiliano Cristiani}
\address{Istituto per le Applicazioni del Calcolo ``M. Picone'', Consiglio Nazionale delle Ricerche, Rome, Italy}
\email{e.cristiani@iac.cnr.it}
\author{Andrea Tosin}
\address{Department of Mathematical Sciences ``G. L. Lagrange'', Politecnico di Torino, Turin, Italy}
\email{andrea.tosin@polito.it}

\begin{document}

\subjclass[2010]{35Q20, 35Q84, 35Q91}

\keywords{Many-particle systems, Boltzmann-type kinetic description, Fokker-Planck equation, multiscale coupling}

\begin{abstract}
In this paper we investigate the possibility of reducing the complexity of a system composed of a large number of interacting agents, whose dynamics feature a symmetry breaking. We consider first order stochastic differential equations describing the behavior of the system at the particle (i.e., Lagrangian) level and we get its continuous (i.e., Eulerian) counterpart via a kinetic description. However, the resulting continuous model alone fails to describe adequately the evolution of the system, due to the loss of granularity which prevents it from reproducing the symmetry breaking of the particle system. By suitably coupling the two models we are able to reduce considerably the necessary number of particles while still keeping the symmetry breaking and some of its large-scale statistical properties. We describe such a multiscale technique in the context of opinion dynamics, where the symmetry breaking is induced by the results of some opinion polls reported by the media.
\end{abstract}

\maketitle

\section{Introduction}
\label{sect:intro}
In this paper we investigate the possibility of reducing the complexity of a system composed of a large number of interacting agents with symmetry breaking. More specifically, we assume that the agents are subject to aggregate stimuli which trigger a loss of symmetry of the particle dynamics.

It is well known that the numerical approximation of multiagent systems becomes rapidly unfeasible when the number of the agents increases, especially when all-to-all interactions are considered. The problem is commonly solved by assuming that the agents are indistinguishable, whereby continuous descriptions are derived in the limit of an infinite number of particles, see e.g.,~\cite{carrillo2010MSSET}. Nevertheless this approach is not always suitable because it loses completely the \emph{particle granularity}, which may instead play a role in the dynamics of the original system especially when the number of particles is large but finite.

Generally speaking, in the following we consider an interacting particle system with the following characteristics:
\begin{itemize}
\item The original full particle model is not directly amenable to numerical computations due to the extremely high number of agents, which implies an excessive computational cost;
\item A particle model with a significantly reduced number of agents departs too much from the actual results;
\item The corresponding averaged continuous model cannot describe adequately the system due to the loss of granularity, which is instead assumed to play a crucial role in the dynamics.
\end{itemize}

We tackle the modeling of such a system by coupling the particle and continuous descriptions in such a way that the latter accounts for most of the agents that are not explicitly tracked in a suitably reduced version of the former and, at the same time, it does not lose the proper contribution of the particle granularity. This method takes inspiration from the multiscale technique first proposed in~\cite{cristiani2011MMS}, then extended in~\cite{cristiani2015JCSMD,cristiani2012CDC,cristiani2014BOOK}, and subsequently applied also in~\cite{cacace2017M2AN}. Such a method is characterized by the fact that a dual microscopic/macroscopic description of a certain particle system is active at all times in the whole domain, with the two scales which continuously complement each other. It is worth stressing that in~\cite{cristiani2011MMS} the microscopic and the macroscopic models are two copies of the same physical system. Instead, one of the novelties of the approach proposed here is that some of the dynamical features of the whole system are confined to either description and do not have a direct counterpart in the other description. Typically this applies to those features mainly responsible for the break of symmetry, which are retained only at the particle level. Most important, we also take full advantage of this multiscale approach to reduce the degrees of freedom of the particle model, which represents the largest source of computational cost.

Such a multiscale approach is investigated here with reference to a specific application in the framework of \emph{opinion dynamics}. We refer the reader to the surveys~\cite{albi2017CHAPTER,aydogdu2017CHAPTER,castellano2009RMP} for an introduction to the topic. We consider a population of interacting individuals who share their opinions about a binary voting choice, such as e.g., ``yes'' or ``no'' in a referendum. Furthermore, the individuals are exposed to a number of opinion polls, whose results can impact on their opinions as well. This feature is mainly responsible for the break of symmetry in the particle dynamics.

Like in the celebrated Hegselmann-Krause model~\cite{hegselmann2002JASSS}, an opinion is described by a continuous variable $w\in\R$. Specifically, we assume $w\in[-1,1]$, so that $\sgn{w}$ expresses the intention of vote (for instance, we may admit that $\sgn{w}=1$ stands for ``yes'' while $\sgn{w}=-1$ stands for ``no'') while $\abs{w}$ gives the degree of conviction.

Regarding the opinion dynamics we make the standard assumption that people tend to \emph{compromise}~\cite{castellano2009RMP,toscani2006CMS}, cf. also~\cite{cucker2007TAC}. This attitude is modulated by the radicalization of their opinion like in~\cite{chowdhury2016CDC}. The interaction with the results of the opinion polls follows a similar principle. However, in this case the individuals do not know the opinions of the interviewed people singularly but only the global prevalence of either voting option in the poll. Thus the results of the opinion polls are known only in aggregate form. A similar feature is found in CODA models~\cite{ceragioli2016PREPRINT,chowdhury2016CDC,martins2008IJMP-C}, where people may see the final action of the others without any access to their real, possibly non-sharp, opinion.

Unlike other models of opinion dynamics, cf. e.g.,~\cite{albi2015CMS,frasca2012NARWA,deffuant2000ACS,hegselmann2002JASSS}, we neglect the \emph{bounded confidence constraint}, namely the fact that the individuals may refuse to interact with people with opinions too far from their own. This is because we want to consider scenarios in which one cannot fully choose who to interact with because one is exposed to partly uncontrollable stimuli: newspapers, broadcasts, reader's comments in blogs on the Internet and alike, where one can accidentally come into contact also with opinions very different from the personal one.

From the technical point of view, we consider a mean-field-type particle description of the opinion dynamics. Next we reduce it to a binary interaction scheme for short time intervals $\Delta{t}>0$, see e.g.,~\cite{albi2015CMS,carrillo2010SIMA,pareschi2013BOOK}, whence we derive a Boltzmann-type kinetic description. Finally we pass to the limit $\Delta{t}\to 0^+$ (\emph{quasi-invariant opinion limit}~\cite{toscani2006CMS}), thereby recovering a Fokker-Planck continuous description of the original mean-field particle model. Nevertheless, the contribution of the opinion polls does not pass to the limit in order to maintain its intrinsically microscopic stocastic features. This originates the necessity to couple the particle and the continuous descriptions, which constitutes the core of the multiscale approach proposed in the paper. It is worth anticipating that such a coupling involves also stochastic terms related to the \emph{self-thinking} of the individuals~\cite{bennaim2005EL}, which have to be handled carefully in order to obtain a consistent multiscale description.

In more detail, the paper is organized as follows. In Section~\ref{sect:micro} we introduce the particle model and we discuss in particular the role of the opinion polls in producing a break of symmetry in the opinion dynamics. In Section~\ref{sect:micro.to.collective} we detail the passage to a continuous model by means of the aforesaid kinetic approach, keeping however a particle description of the opinion polls. In Section~\ref{sect:hybrid.micro.multiscale} we propose an hybridization of a reduced version of the particle model with the continuous model, thereby creating a coupled multiscale model. Such an hybridization aims, on one hand, at replacing most of the particles with a continuous description by means of their probability distribution function and, on the other hand, at retaining a proper (small) number of them embedded in the continuous model in order to simulate reliably the opinion polls. In Section~\ref{sect:num_approx} we describe the numerical approximation of the multiscale model and in Section~\ref{sect:simulations} we perform several numerical tests to show the performances of the multiscale coupling in reducing the complexity of the full particle description. Finally, in Section~\ref{sect:conclusions} we draw some conclusions and briefly sketch research perspectives.

\section{Particle-based microscopic modeling}
\label{sect:micro}
\subsection{Basic opinion dynamics}
\label{sect:opinion.dynamics}
We consider a system of $N$ agents characterized by their opinion $w_k$, $k=1,\,\dots,\,N$, which has to be understood as a function of time: $w_k=w_k(t):[0,\,T]\to [-1,\,1]$, where $T>0$ is some fixed final instant, possibly $T=+\infty$. At each time, $w_k$ is assumed to belong to the bounded interval $[-1,\,1]$ with the convention that $\sgn{w_k}=\pm 1$ identifies the binary option (e.g., ``yes'' or ``no'') that the $k$th voter supports while $\abs{w_k}\in [0,\,1]$ expresses his/her conviction for that option.

Taking inspiration from~\cite{brugna2015PRE,toscani2006CMS}, we assume that the opinion $w_k$ changes in time because of two basic mechanisms:
\begin{itemize}
\item the interaction with the opinions of the other agents;
\item the \emph{self-thinking} of the $k$th agent, which we interpret as a random walk in the space of the opinions, see also~\cite{bennaim2005EL}.
\end{itemize}

In view of these assumptions, we write:
\begin{equation}
	dw_k=a(w_k)\left(\frac{1}{N-1}\sum_{h\ne k}I(w_k,\,w_h)dt+\sqrt{2\mu}dB^k_t\right), \qquad k=1,\,\dots,\,N,
	\label{eq:micro}
\end{equation}
where:
\begin{itemize}
\item $a:[-1,\,1]\to\R_+$ is a function expressing the \emph{propensity to change opinion}. We assume
\begin{equation}
	a(w)=1-\abs{w}^\delta, \qquad \delta>0,
	\label{eq:a}
\end{equation}
the underlying idea being that the more radical the opinion the lower the propensity to change it;
\item $I:[-1,\,1]^2\to\R$ is a function modeling the interaction between two opinions. According to the main trend in the reference literature~\cite{krause2000CHAPTER,toscani2006CMS}, $I$ is usually proportional to the relative opinion of the interacting agents:
\begin{equation}
	I(w,\,v)=\alpha(v-w), \qquad \alpha>0,
	\label{eq:I}
\end{equation}
meaning that interactions tend to lead mostly to a \emph{compromise}; namely, the post-interaction opinions are closer than the pre-interaction ones.
\item $B^k_t$ is, for every $k=1,\,\dots,\,N$, a standard Brownian motion (Wiener process) with independent Gaussian increments and $\mu>0$ is a constant. In particular, $B^k_{t+s}-B^k_t$, $s\geq 0$, has normal distribution $\mathcal{N}(0,\,s)$ with zero mean and variance $s$, while $B^k_t$, $B^h_t$ are, for $k\ne h$ and each $t\geq 0$, independent and identically distributed.
\end{itemize}

\subsection{The opinion poll}
\label{sect:opinion.poll}
The basic opinion dynamics modeled by~\eqref{eq:micro} may be influenced, in case of political campaigns, by the results of several \emph{opinion polls} performed before the vote to test the feelings of the voters.

A poll consists in interviewing a representative sample of voters about their current voting intentions and then in processing statistically their answers to simulate the distribution of the opinions in the whole population. The size of the sample which guarantees a significant confidence level is in general a function of the total number $N$ of the individuals, which tends to saturate when $N$ grows~\cite{wilks1940POQ}. In order to get rid of the uncertainty linked to the proper size of the sample, in our model we are going to assume that \emph{all} voters are interviewed in each poll, hence the results of the polls can be regarded as exact by definition.

Let $\{T_i\}_{i=1}^{N_P}$ be a set of times, $0<T_i<T_{i+1}<T$ all $i$, at which $N_P\geq 1$ polls are performed. At each poll, the percentages of the interviewed individuals who are in favor of either option are computed, that is respectively:
\begin{align*}
	P_i^+ &:= \frac{\card\{k\in\{1,\,\dots,\,N\}\,:\,w_k(T_i)>W^0\}}{N}, \\
	P_i^- &:= \frac{\card\{k\in\{1,\,\dots,\,N\}\,:\,w_k(T_i)<-W^0\}}{N},
\end{align*}
$W^0\in [0,\,1)$ being a ``null threshold'' such that if $\abs{w_k}\leq W^0$ then the $k$th individual is regarded as indecisive and his/her opinion is not counted in either pool. Out of them, we define the gap
$$ P_i:=P_i^+-P_i^-\in [-1,\,1], $$
which we assume to bias the opinions of the voters over a certain (usually short) time period immediately after the release of the poll result. We model this effect by assuming that $P_i$ perturbs the null average of the self-thinking in~\eqref{eq:micro}:
\begin{equation}
	dw_k=a(w_k)\left(\frac{1}{N-1}\sum_{h\ne k}I(w_k,\,w_h)dt+\sqrt{2\mu}d\tilde{B}^k_t\right), \qquad k=1,\,\dots,\,N,
	\label{eq:micro.poll}
\end{equation}
where $\tilde{B}^k_t$ is a new noise term for the $k$th individual with independent Gaussian increments, such that $\tilde{B}^k_{t+s}-\tilde{B}^k_t$, $s\geq 0$, has normal distribution with
\begin{equation}
	\avefences{\tilde{B}^k_{t+s}-\tilde{B}^k_t}=\beta\int_t^{t+s}b(\tau,\,\{P_i\}_{i=1}^{N_P},\,w_k)\,d\tau,
		\qquad \operatorname{Var}(\tilde{B}^k_{t+s}-\tilde{B}^k_t)=s,
	\label{eq:Btilde.ave.var}
\end{equation}
where $\avefences{\cdot}$ is the expectation and $\beta>0$ is a constant. The function $b$ models how the self-thinking of the $k$th individual is biased on average over time by the results of the various opinion polls and possibly also by his/her current opinion. Notice that if $b$ is independent of $w_k$ then the noise $\tilde{B}^k_t$ is actually the same for all individuals, i.e., it is independent of $k$ as well. Several choices are conceivable for $b$. We consider in particular the following two options:
\begin{subequations}
\begin{gather}
	b(t,\,\{P_i\}_{i=1}^{N_P},\,w_k):=\sum_{i=1}^{N_P}\abs{P_i}^\nu(\sgn{P_i}-w_k)\eta(t-T_i) \label{eq:b.sgnDP-wk} \\
	b(t,\,\{P_i\}_{i=1}^{N_P},\,w_k):=\sum_{i=1}^{N_P}\abs{P_i}^\nu\sgn{P_i}\,\eta(t-T_i) \label{eq:b.DP},
\end{gather}
\end{subequations}
with $\nu>0$, where in both cases $\eta:\R\to\R_+$ is a function giving the duration in time of the influence of a poll. We assume $\eta(s)=0$ for $s<0$, so that a poll has no influence before the instant at which it is performed, and $\eta\to 0$ sufficiently fast for $s\to +\infty$. For instance:
\begin{equation}
	\eta(s):=\frac{1}{\Delta{T}}\mathbbm{1}_{[0,\,\Delta{T}]}(s)
	\label{eq:eta.chi}
\end{equation}
or
$$ \eta(s):=\frac{1}{\Delta{T}}e^{-s/\Delta{T}}\mathbbm{1}_{[0,\,+\infty)}(s) $$
for a suitable choice of the parameter $\Delta{T}$ which defines the characteristic duration of the effect of a poll. In the limit $\Delta{T}\to 0^+$ we obtain $\eta(s)=\delta_0(s)$, the Dirac delta centered at the origin, indicating an \emph{impulsive} poll.

The function~\eqref{eq:b.sgnDP-wk} expresses a drag of $w_k$ toward $\sgn{P_i}$, understood as representative of the prevailing binary option at the $i$th poll, modulated by $\abs{P_i}^\nu$, which gives the strength of the $i$th poll. The function~\eqref{eq:b.DP} expresses instead a bias of the average self-thinking which depends only on the result of the polls and acts therefore equally on all the voters. Notice that for $t\in\R_+\setminus\cup_{i=1}^{N_P}[T_i,\,T_i+\Delta{T}]$, i.e., in the time instants in which no poll has effect, both~\eqref{eq:b.sgnDP-wk} and~\eqref{eq:b.DP} imply $b\equiv 0$. Hence, if $\eta$ is the function in~\eqref{eq:eta.chi}, in those instants $\tilde{B}^k_t$ reduces to the standard Brownian motion like in~\eqref{eq:micro}.

\subsection{The result of the vote}
\label{sect:result.vote.micro}
At time $t=T$ the final vote takes place and its result can be assessed similarly to each poll. Specifically, we define the percentages of individuals who at last voted for either option:
$$ V^\pm:=\frac{\card\{k\in\{1,\,\dots,\,N\}\,:\,\sgn{w_k(T)}=\pm 1\}}{N} $$
and the corresponding gap
$$ V:=V^+-V^-\in [-1,\,1]. $$

\section{From the particle to a continuous description}
\label{sect:micro.to.collective}
\subsection{Binary interaction dynamics}
\label{sect:binary.interactions}
Taking inspiration from~\cite{albi2015CMS,toscani2006CMS}, we consider the dynamics~\eqref{eq:micro} between two individuals $k$ and $h$ taking place in a short time $\Delta{t}>0$. If we denote by $w_k^\ast=w_k(t+\Delta{t})$ the post-interaction opinion of the individual $k$ then we deduce from~\eqref{eq:micro.poll}
\begin{equation}
	w_k^\ast=w_k+a(w_k)\left(I(w_k,\,w_h)\Delta{t}+\sqrt{2\mu}\Delta{\tilde{B}}^k_t\right)+o(\Delta{t}).
	\label{eq:micro.poll.approx}
\end{equation}
Notice that, according to~\eqref{eq:Btilde.ave.var} together with a first order approximation of the integral in time contained in $avefences{\tilde{B}^k_{t+\Delta{t}}-\tilde{B}^k_t}$, it results
$$ \Delta{\tilde{B}}^k_t=\tilde{B}^k_{t+\Delta{t}}-\tilde{B}^k_t
	\sim\mathcal{N}\left(\beta b(t,\,\{P_i\}_{i=1}^{N_P},\,w_k)\Delta{t}+o(\Delta{t}),\,\Delta{t}\right), $$
which can be normalized by defining the random variable
$$ Z:=\frac{\Delta{\tilde{B}}^k_t-\beta b(t,\,\{P_i\}_{i=1}^{N_P},\,w_k)\Delta{t}+o(\Delta{t})}{\sqrt{\Delta{t}}}
	\sim\mathcal{N}(0,\,1). $$
Taking this into account and assuming that in a large population the individuals are mostly indistinguishable,~\eqref{eq:micro.poll.approx} can be finally abstracted for two generic agents with pre-interaction opinions $w,\,v$ as
\begin{equation}
	w^\ast=w+a(w)\left[\left(I(w,\,v)+\sqrt{2\mu}\beta b(t,\,\{P_i\}_{i=1}^{N_P},\,w)\right)\Delta{t}
		+\sqrt{2\mu\Delta{t}}Z\right],
	\label{eq:binary}
\end{equation}
where we have suppressed the term $o(\Delta{t})$ while still enforcing the equality.

The binary interaction rule~\eqref{eq:binary} can be encoded in a Boltzmann-type kinetic description of the system. To this end, we introduce the reference time scale of the binary interactions:
\begin{equation}
	\tau:=\frac{t}{\Delta{t}},
	\label{eq:t-tau}
\end{equation}
where the characteristic time of a binary interaction becomes $O(1)$, and we consider the kinetic distribution function $g=g(\tau,\,w)$ such that $g(\tau,\,w)dw$ is the probability that at time $\tau>0$ a generic individual has an opinion in $[w,\,w+dw]$. Consistently with~\eqref{eq:binary}, the function $g$ satisfies the following Boltzmann-type equation, cf.~\cite{pareschi2013BOOK}:
\begin{equation}
	\frac{d}{d\tau}\int_{-1}^{1}\varphi(w)g(\tau,\,w)\,dw=
		\avefences*{\int_{-1}^{1}\int_{-1}^{1}\left(\varphi(w^\ast)-\varphi(w)\right)g(\tau,\,w)g(\tau,\,v)\,dw\,dv}
	\label{eq:boltzmann}
\end{equation}
for every test function $\varphi:[-1,\,1]\to\R$. Here $\avefences{\cdot}$ denotes the expectation with respect to the (independent) random variable $Z$ appearing in~\eqref{eq:binary}.

\begin{remark}
For the moment we deliberately neglect the question of whether $w^\ast$ defined by~\eqref{eq:binary} belongs to the interval $[-1,\,1]$. We will come back to this issue at the end of the next section, cf. Remark~\ref{rem:mass_cons.fokker-planck}, after obtaining the definitive continuous counterpart of model~\eqref{eq:micro.poll}.
\end{remark}

\subsection{Fokker-Planck asymptotic analysis}
\label{sect:fokker-planck.analysis}
Equation~\eqref{eq:boltzmann} is a high-resolution one in time, because it describes the evolution of $g$ subject to binary interactions on the short $\tau$-scale. Actually, such a detail is not needed to depict the collective trends of the system, also in view of the mean-field nature of the interactions in the original particle model~\eqref{eq:micro}. With this in mind, equation~\eqref{eq:boltzmann} can be profitably used to extract the principal part of the interactions directly on the coarser $t$-scale.

To this purpose, invoking~\eqref{eq:t-tau} we scale the kinetic distribution function as
$$ f(t,\,w):=g(t/\Delta{t},\,w). $$
Clearly, $\int_{-1}^{1}f(t,\,w)\,dw=1$ while $\partial_t f=\frac{1}{\Delta{t}}\partial_\tau g$. By evaluating the right-hand side of~\eqref{eq:boltzmann} for $\tau=t/\Delta{t}$ it follows then that $f$ satisfies the equation
\begin{equation}
	\frac{d}{dt}\int_{-1}^{1}\varphi(w)f(t,\,w)\,dw=
		\frac{1}{\Delta{t}}\avefences*{\int_{-1}^{1}\int_{-1}^{1}\left(\varphi(w^\ast)-\varphi(w)\right)f(t,\,w)f(t,\,v)\,dw\,dv}.
	\label{eq:boltzmann.f}
\end{equation}

We observe from~\eqref{eq:binary} that $w^\ast-w=O(\sqrt{\Delta{t}})$. Since $\Delta{t}$ is small, if we take $\varphi$ sufficiently smooth, with moreover $\varphi(\pm 1)=\varphi'(\pm 1)=0$, we can expand
$$ \varphi(w^\ast)-\varphi(w)=\varphi'(w)(w^\ast-w)+\frac{1}{2}\varphi''(w)(w^\ast-w)^2
	+\frac{1}{6}\varphi'''(\bar{w})(w^\ast-w)^3, $$
where $\bar{w}$ is a point such that $\min\{w,\,w^\ast\}<\bar{w}<\max\{w,\,w^\ast\}$. Substituting the expression of $w^\ast$ and plugging into~\eqref{eq:boltzmann.f} we obtain
\begin{align}
	\begin{aligned}[b]
		&\frac{d}{dt}\int_{-1}^{1}\varphi(w)f(t,\,w)\,dw \\
		&= \int_{-1}^{1}\int_{-1}^{1}\varphi'(w)a(w)\left(I(w,\,v)+\sqrt{2\mu}\beta b(t,\,\{P_i\}_{i=1}^{N_P},\,w)\right)f(t,\,w)f(t,\,v)\,dw\,dv \\
		&\phantom{=} +\mu\int_{-1}^{1}\varphi''(w)a^2(w)f(t,\,w)\,dw+R(\Delta{t}),
	\end{aligned}
	\label{eq:boltzmann.f-remainder}
\end{align}
the remainder $R$ being
\begin{align*}
	R(\Delta{t}) &:= \frac{\Delta{t}}{2}\int_{-1}^{1}\int_{-1}^{1}\varphi''(w)a^2(w)\left(I+\sqrt{2\mu}\beta b\right)^2f(t,\,w)f(t,\,v)\,dw\,dv \\
	&\phantom{:=} +\frac{\sqrt{\Delta{t}}}{6}\avefences*{\int_{-1}^{1}\int_{-1}^{1}\varphi'''(\bar{w})a^3(w)H^3f(t,\,w)f(t,\,v)\,dw\,dv}
\end{align*}
where we have denoted $H:=\sqrt{\Delta{t}}\left(I+\sqrt{2\mu}\beta b\right)+\sqrt{2\mu}Z$ for brevity. Since
$$ \abs{R(\Delta{t})}\leq\frac{\Delta{t}}{2}C^2\norm{a}^2\norm{\varphi''}
	+\Delta{t}\norm{a}^3\left(\frac{\Delta{t}}{6}C^2+\mu\right)C\norm{\varphi'''}, $$
where $C:=\norm{I}+\sqrt{2\mu}\beta\norm{b}$, we deduce $R(\Delta{t})\to 0$ for $\Delta{t}\to 0^+$. In such a limit we obtain from~\eqref{eq:boltzmann.f-remainder}
\begin{align}
	\begin{aligned}[b]
		&\frac{d}{dt}\int_{-1}^{1}\varphi(w)f(t,\,w)\,dw \\
		&= \int_{-1}^{1}\int_{-1}^{1}\varphi'(w)a(w)\left(I(w,\,v)+\sqrt{2\mu}\beta b(t,\,\{P_i\}_{i=1}^{N_P},\,w)\right)f(t,\,w)f(t,\,v)\,dw\,dv \\
		&\phantom{=} +\mu\int_{-1}^{1}\varphi''(w)a^2(w)f(t,\,w)\,dw,
	\end{aligned}
	\label{eq:fokker-planck.weak}
\end{align}
which, since $\varphi$, $\varphi'$ vanish for $w=\pm 1$, is a weak form of
\begin{equation}
	\partial_tf+\partial_w(K[f]f)=\mu\partial^2_w(a^2(w)f),
	\label{eq:fokker-planck}
\end{equation}
where
\begin{equation}
	K[f]=K[f](t,\,w):=a(w)\left(\int_{-1}^{1}I(w,\,v)f(t,\,v)\,dv+\sqrt{2\mu}\beta b(t,\,\{P_i\}_{i=1}^{N_P},\,w)\right)
	\label{eq:K}
\end{equation}
is the rate of change of the opinion $w$ due to the interactions with other agents and with the results of the opinion polls.

Equation~\eqref{eq:fokker-planck} is the Fokker-Planck approximation of the Boltzmann equation \eqref{eq:boltzmann} on the coarser $t$-scale, where the collective mean-field outcome of the binary interactions is directly observable. As such, it is directly comparable with~\eqref{eq:micro.poll} when the total number $N$ of the agents is sufficiently large.

\begin{remark}[Mass conservation in the Fokker-Planck equation] 	\label{rem:mass_cons.fokker-planck}
Integrating \eqref{eq:fokker-planck} with respect to $w\in [-1,\,1]$ we discover
$$ \frac{d}{dt}\int_{-1}^{1}f(t,\,w)\,dw+\Bigl[K[f]f\Bigr]_{-1}^{1}=\mu\Bigl[\partial_w(a^2(w)f)\Bigr]_{-1}^{1}, $$
whence, since $\partial_w(a^2(w)f)=a(w)\left(2a'(w)f+a(w)\partial_wf\right)$ with $a'\in L^\infty(-1,\,1)$ for $a(w)$ given by~\eqref{eq:a} and recalling that $a(\pm 1)=0$,
$$ \frac{d}{dt}\int_{-1}^{1}f(t,\,w)\,dw=0. $$
This indicates that if the initial distribution $f_0(w):=f(0,\,w)$ is such that $\supp{f_0}\subseteq [-1,\,1]$ then $\supp{f(t,\,\cdot)}\subseteq [-1,\,1]$ for all $t>0$. Therefore, the bounds $-1\leq w\leq 1$ are almost surely never violated.
\end{remark}

\section{Hybrid particle model and multiscale coupling}
\label{sect:hybrid.micro.multiscale}
Model~\eqref{eq:micro.poll} consists of $N$ coupled ordinary differential equations. As already pointed out in Section~\ref{sect:micro.to.collective}, such a description may not be satisfactory when the total number $N$ of the agents gets large.

The corresponding continuous description provided by the Fokker-Planck equation~\eqref{eq:fokker-planck} solves only partly this issue. In fact the advection term $b(t,\,\{P_i\}_{i=1}^{N_P},\,w)$ in $K[f]$ still depends on the poll gaps $P_i$, which are intrinsically microscopic quantities. Since the interaction with the polls is the main factor triggering the break of symmetry in the particle dynamics, it is quite natural, and possibly also necessary, to keep its description at the particle level. This is especially true if the number $N$ is large but finite for then, as recalled in the Introduction, the particle granularity can possibly play a non-negligible role in the overall dynamics. Therefore solving directly~\eqref{eq:fokker-planck} is even less feasible than addressing system~\eqref{eq:micro.poll} alone, in fact the Fokker-Planck equation is simply an additional equation to be coupled to the entire particle model.

In order to bypass these difficulties of the theory, we propose a \emph{multiscale} approach based on the \emph{hybridization} of a \emph{reduced} particle model, i.e. a model of type~\eqref{eq:micro.poll} with $N_\ast\ll N$ agents, with the Fokker-Planck equation~\eqref{eq:fokker-planck}. The latter is meant to replace the remaining larger part of the agents that are not represented individually.

As anticipated in the Introduction, this method takes inspiration from the multiscale approach originally developed in~\cite{cristiani2011MMS,cristiani2014BOOK} for an integrated micro/macroscopic representation of a particle system without noise. Here we adapt the technique by including the diffusive contribution like in~\cite{cristiani2015JCSMD} and we evolve the whole approach according to the ideas just outlined.

To begin with, from the computations made in Section~\ref{sect:micro.to.collective} we observe that we can write
$$ d\tilde{B}^k_t=\sqrt{2\mu}\beta b(t,\,\{P_i\}_{i=1}^{N_P},\,w_k)dt+dB^k_t, $$
thus~\eqref{eq:micro.poll} can be restated for the reduced population as
\begin{align}
	\begin{aligned}[b]
		dw_k=a(w_k)\Biggl[&\left(\frac{1}{N_\ast-1}\sum_{h=1}^{N^\ast}I(w_k,\,w_h)
			+\sqrt{2\mu}\beta b(t,\,\{P_i\}_{i=1}^{N_P},\,w_k)\right)dt \\
		&+\sqrt{2\mu}dB^k_t\Biggr]
	\end{aligned}
	\label{eq:reduced.micro}
\end{align}
with $k=1,\,\dots,\,N_\ast$ and
\begin{align*}
	P_i &=\frac{\card{\{k\in\{1,\,\dots,\,N_\ast\}\,:\,w_k(T_i)>W^0\}}}{N_\ast} \\
	&\phantom{=} -\frac{\card{\{k\in\{1,\,\dots,\,N_\ast\}\,:\,w_k(T_i)<-W^0\}}}{N_\ast}.
\end{align*}
In the sum at the right-hand side of~\eqref{eq:reduced.micro} we have taken into account in particular that $I(w_k,\,w_k)=0$ according to~\eqref{eq:I}.

In order to hybridize this model with the Fokker-Planck equation~\eqref{eq:fokker-planck}, we first proceed to rewrite the latter in flux form so as to put in evidence a transport velocity in the space of the opinions. We have:
$$ \partial_tf+\partial_w\left[\left(K[f]-\mu\partial_wa^2(w)-\mu a^2(w)\frac{\partial_wf}{f}\right)f\right]=0, $$
whence we identify the diffusive contribution to the rate of change of $f$:
\begin{equation}
	D[f](t,\,w):=\mu\left(\partial_wa^2(w)+a^2(w)\frac{\partial_wf}{f}\right)=\mu a^2(w)\partial_w\log{(a^2(w)f)}.
	\label{eq:D}
\end{equation}
Thus we are led to consider the Fokker-Planck equation in the conservative form
\begin{equation}
	\partial_tf+\partial_w((K[f]-D[f])f)=0.
	\label{eq:fokker-planck.flux}
\end{equation}

In order to embed the reduced microscopic model into this equation the idea is to hybridize the drift and self-thinking terms in~\eqref{eq:reduced.micro} with $K[f]$ and $-D[f]$, respectively, from~\eqref{eq:fokker-planck.flux}. Specifically, we consider a convex linear combination by means of a parameter $\theta\in [0,\,1]$:
\begin{align}
	\begin{aligned}[b]
		dw_k &= \Biggl[\theta a(w_k)\biggl(\frac{1}{N_\ast-1}\sum_{h=1}^{N_\ast}I(w_k,\,w_h)
			+\sqrt{2\mu}\beta b(t,\,\{P_i\}_{i=1}^{N_P},\,w_k)\biggl) \\
		&\phantom{=} +(1-\theta)K[f](t,\,w_k)\Biggr]dt \\
		&\phantom{=} +\Theta\sqrt{2\mu}a(w_k)dB^k_t+(1-\Theta)\left(-D[f](t,\,w_k)\right)dt,
	\end{aligned}
	\label{eq:hybrid.micro}
\end{align}
for $k=1,\,\dots,\,N_\ast$, where $\Theta\sim\text{Bernoulli}(\theta)$ is an independent random variable which we use to hybridize the particle and continuous diffusive contributions. In~Appendix~\ref{app:coupling.diffusion} we prove that this stochastic hybridization is indeed necessary for a consistent coupling of the two diffusive contributions, see also~\cite{cristiani2015JCSMD}.

\begin{figure}[!t]
\centering
\includegraphics[width=\textwidth]{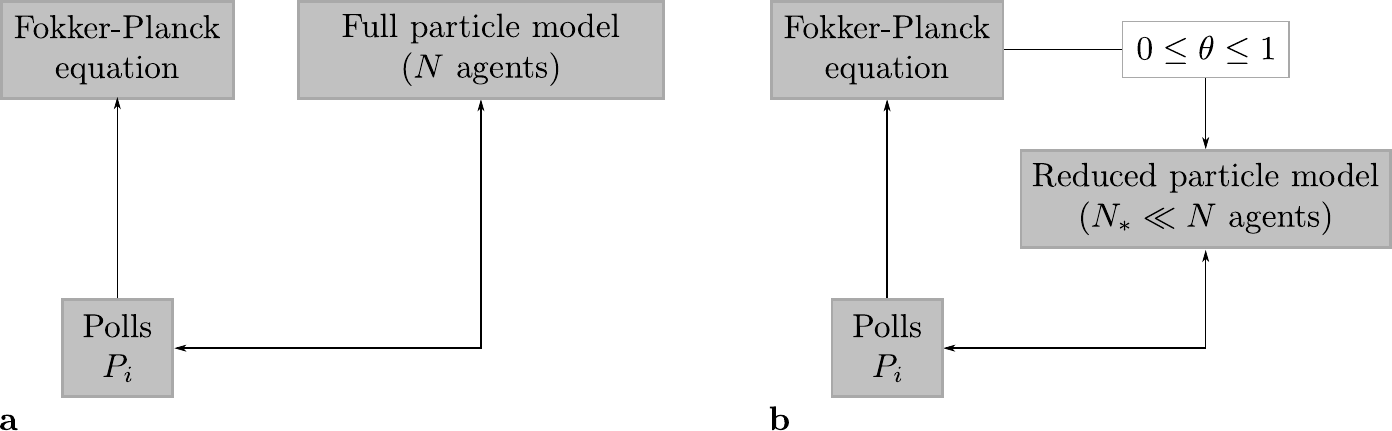}
\caption{\textbf{a}. Solving directly the Fokker-Planck equation~\eqref{eq:fokker-planck} would require to solve in parallel the full particle model~\eqref{eq:micro.poll} to obtain the poll gaps $P_i$. -- \textbf{b}. The multiscale coupling between~\eqref{eq:fokker-planck}-\eqref{eq:K} and~\eqref{eq:hybrid.micro} allows one to use a strongly reduced number of agents to compute the $P_i$'s.}
\label{fig:multiscale.diagram}
\end{figure}

In the hybrid equation~\eqref{eq:hybrid.micro} the $N_\ast$ agents of the reduced particle model~\eqref{eq:reduced.micro} are complemented with a continuous representation of the $N-N_\ast\gg N_\ast$ agents not modeled individually. On one hand, this allows one to compute correctly the results $\{P_i\}_{i=1}^{N_P}$ of the polls at the particle level, which would not be possible using~\eqref{eq:reduced.micro} alone because of the lack of most of the agents of the population. On the other hand, the correct computation of the $P_i$'s is crucial to obtain from~\eqref{eq:fokker-planck}-\eqref{eq:K} a reliable evolution of $f$ to be used in~\eqref{eq:hybrid.micro}. Figure~\ref{fig:multiscale.diagram} illustrates the conceptual difference between this multiscale approach and the unsatisfactory direct approach to the solution of~\eqref{eq:fokker-planck} mentioned at the beginning of this section.

In the following we will refer to~\eqref{eq:hybrid.micro} as the \emph{hybrid particle model} (or the \emph{hybrid particle equation}) and to the coupled model~\eqref{eq:fokker-planck.flux}-\eqref{eq:hybrid.micro} as the \emph{multiscale model}.

\section{Numerical approximation of the equations}
\label{sect:num_approx}
We now describe the numerical approximation of the multiscale model~\eqref{eq:fokker-planck.flux}-\eqref{eq:hybrid.micro}.

We approximate~\eqref{eq:fokker-planck.flux} via the standard first order upwind scheme after discretizing the space of the opinions $[-1,\,1]$ in $J>0$ cells
\begin{equation}
	E_j:=\left[W_j-\frac{\Delta{w}}{2},\,W_j+\frac{\Delta{w}}{2}\right)
	\label{eq:Ej}
\end{equation}
with $\Delta{w}:=\frac{2}{J-1}$, $W_j:=-1+(j-1)\Delta{w}$, $j=1,\,\dots,\,J$, and the time interval $[0,\,T]$ in $M>0$ instants $t^m:=(m-1)\Delta{t}$, $m=1,\,\dots,\,M$, with $\Delta{t}:=\frac{T}{M-1}$. We refer the reader to~\cite{pareschi2017PREPRINT} for higher order numerical schemes for Fokker-Planck equations.

On the other hand, the hybrid particle equation~\eqref{eq:hybrid.micro} can be rewritten as
\begin{align}
	\begin{aligned}[b]
		dw_k &= \Biggl[\theta a(w_k)\left(\frac{1}{N_\ast-1}\sum_{h=1}^{N_\ast}I(w_k,\,w_h)+\sqrt{2\mu}\beta b(t,\,\{P_i\}_{i=1}^{N_P},\,w_k)\right) \\
		&\phantom{=} +(1-\theta)K[f](t,\,w_k)\Biggr]dt+
			\begin{cases}
				\sqrt{2\mu}a(w_k)dB^k_t & \text{with probability } \theta \\[1mm]
				-D[f](t,\,w_k)dt & \text{with probability } 1-\theta.
			\end{cases}
	\end{aligned}
	\label{eq:hybrid.micro.numerics}
\end{align}
We discretize~\eqref{eq:hybrid.micro.numerics} on the same grid $\{t^m\}_{m=1}^{M}$ previously introduced via the strongly consistent Euler scheme. For $\theta=1$ we obtain also the discretization of the reduced particle model~\eqref{eq:reduced.micro}.

In the subsequent numerical simulations we will invariably consider the form~\eqref{eq:a} of the function $a$ with $\delta=1$. With this choice it results
$$ D[f](t,\,w)=\mu\left(2(w-\sgn{w})+(1-\abs{w})^2\frac{\partial_wf}{f}\right). $$

\section{Numerical tests}
\label{sect:simulations}
With the functions $a$, $I$ given by~\eqref{eq:a},~\eqref{eq:I}, the parameters accounting for the strength of the three main factors in~\eqref{eq:fokker-planck.flux}-\eqref{eq:hybrid.micro.numerics}, i.e., the compromise, the polls, and the self-thinking, are
$$ \alpha\ \text{(compromise)}, \qquad p:=\sqrt{2\mu}\beta\ \text{(polls)}, \qquad \mu\ \text{(self-thinking)}, $$
which we will vary in the next simulations together with a few other ones, such as e.g., $N$ and $N_\ast$. 

We assume an interaction with the results of the polls of type~\eqref{eq:b.sgnDP-wk}. In particular, we consider different scenarios with $N_P=0,\,1,\,2$ polls. Initially the opinions $\{w_k\}_{k=1}^{N_\ast}$ are either randomly sampled from the uniform continuous density $f_0(w)=\frac{1}{2}\mathbbm{1}_{[-1,\,1]}(w)$ or evenly distributed in $[-1,\,1]$. Notice that the latter is a quite artificial initial condition for the particle model, which however can be expected to provide in general a better agreement with the output of the continuous model with initial condition $f_0$. In fact it attenuates the impact of the granularity, which arguably is mainly influential in less ordered particle configurations.

All model parameters are summarized in Table~\ref{tab:param}.

\begin{table}[!t]
\caption{Model parameters for the numerical tests of Section~\ref{sect:simulations}}
\centering
\begin{tabular}{|c|c|c|c|c|}
\hline
Parameter & $N_P=0,\,1$ & $N_P=2$ & Explanation & Reference \\
\hline
$T$ & $0.2$ & $0.8$ & Final time of the simulation & Section~\ref{sect:opinion.dynamics} \\
$\alpha$ & $8$ & $8$ & Strength of the compromise & \eqref{eq:I} \\
$p$ & $0.2$ & $0.2$ & Strength of the polls & Section~\ref{sect:simulations} \\
$\mu$ & $0.5$ & $0.5$ & Strength of the self-thinking & Section~\ref{sect:opinion.dynamics} \\
$\delta$ & $1$ & $1$ & Exponent in the function $a$ & \eqref{eq:a} \\
$\nu$ & $\frac{1}{2}$ & $\frac{1}{2}$ & Exponent in the function $b$ & \eqref{eq:b.sgnDP-wk} \\
$W^0$ & $0.1$ & $0.1$ & Threshold of indecisiveness & Section~\ref{sect:opinion.poll} \\
$T_1$ & $0.1$ & $0.1$ & Time of the first poll & Section~\ref{sect:opinion.poll} \\
$T_2$ & -- & $0.5$ & Time of the second poll & Section~\ref{sect:opinion.poll} \\
$\Delta{T}$ & $0.05$ & $0.05$ & Duration of a poll influence & \eqref{eq:eta.chi} \\
$J$ & $51$ & $51$ & Grid nodes in $[-1,\,1]$ & Section~\ref{sect:num_approx} \\
$M$ & $200$ & $800$ & Grid nodes in $[0,\,T]$ & Section~\ref{sect:num_approx} \\
\hline
\end{tabular}
\label{tab:param}
\end{table}

\subsection{Scale comparison}
Let us first illustrate the effect of the single terms of the model (compromise, poll, self-thinking) without hybridization, i.e., $\theta=1$ in~\eqref{eq:hybrid.micro.numerics}. In the following figures we show the trajectories $t\mapsto w_k(t)$, $k=1,\,\dots,\,N$, of the particle opinions for $t\in [0,\,T]$ and the final density $f(T,\,w)$ of the continuous model. The initial opinions $w_k(0)$ are taken evenly spaced in $[-1,\,1]$.

\begin{figure}[!t]
\centering
\includegraphics[width=\figscale\textwidth]{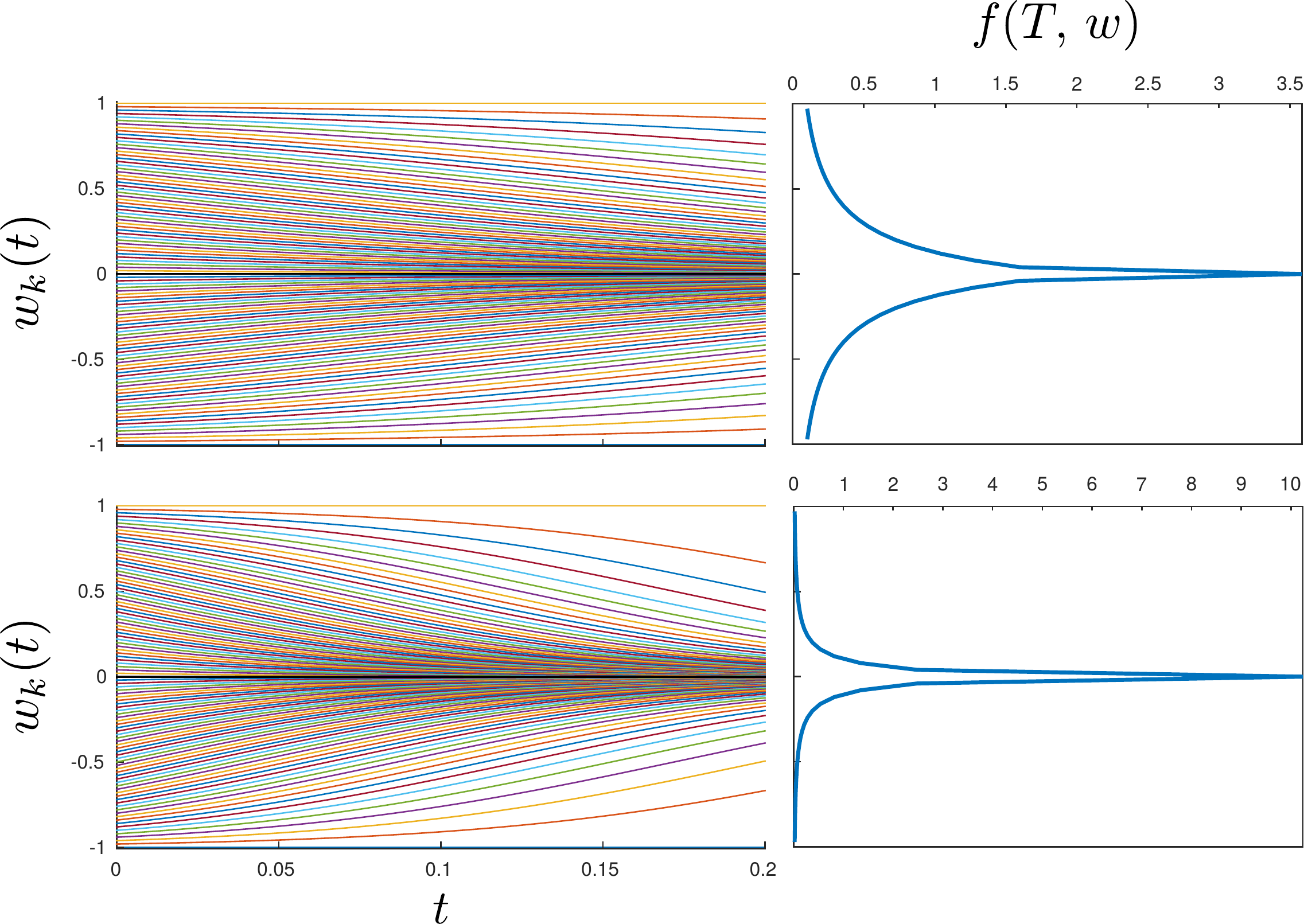}
\caption{Effect of the compromise alone ($p=\mu=0$). Top: the coefficient $\alpha$ is like in Table~\ref{tab:param}. Bottom: the coefficient $\alpha$ is doubled.}
\label{fig:time_trend-only_compromise}
\end{figure}

Figure~\ref{fig:time_trend-only_compromise} shows the result of the simulation when only compromise among the individuals is included in the model ($p=\mu=0$). Due to the symmetry of the initial data, the interactions tend to move the agents toward a common opinion corresponding to the average of the initial distribution ($0$ in this case). By increasing $\alpha$ we observe a faster convergence.

\begin{figure}[!t]
\centering
\includegraphics[width=\figscale\textwidth]{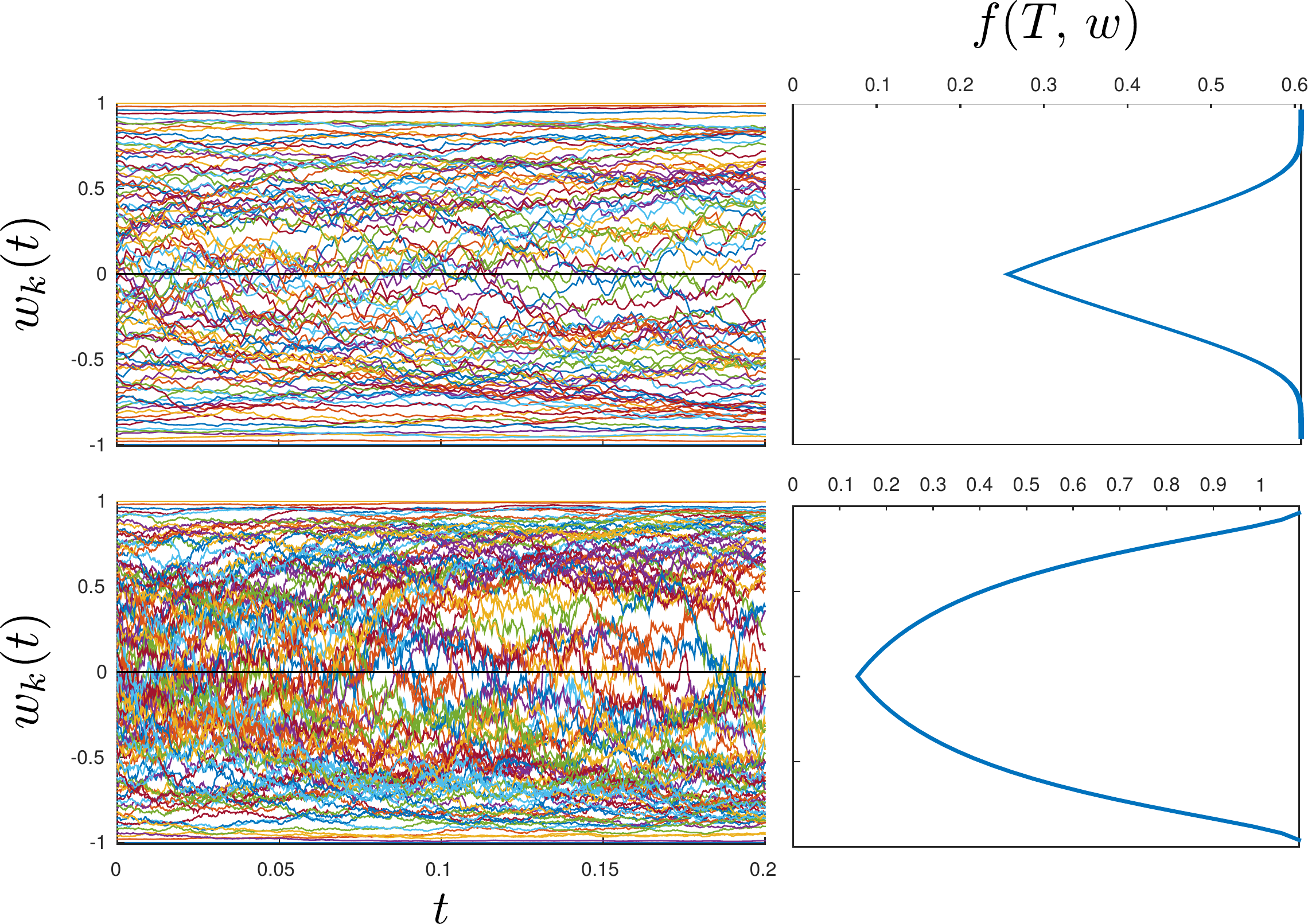}
\caption{Effect of the self-thinking alone ($\alpha=p=0$). Top: the coefficient $\mu$ is like in Table~\ref{tab:param}. Bottom: the coefficient $\mu$ is doubled.}
\label{fig:time_trend-only_self-thinking}
\end{figure}

Figure~\ref{fig:time_trend-only_self-thinking} shows the result of the simulation when only the self-thinking is included in the model ($\alpha=p=0$). Opinions move away from the center and concentrate at the boundaries. By increasing $\mu$ the radicalization is faster. Also in this case the symmetry of the distribution is preserved in time.

\begin{figure}[!t]
\centering
\includegraphics[width=\figscale\textwidth]{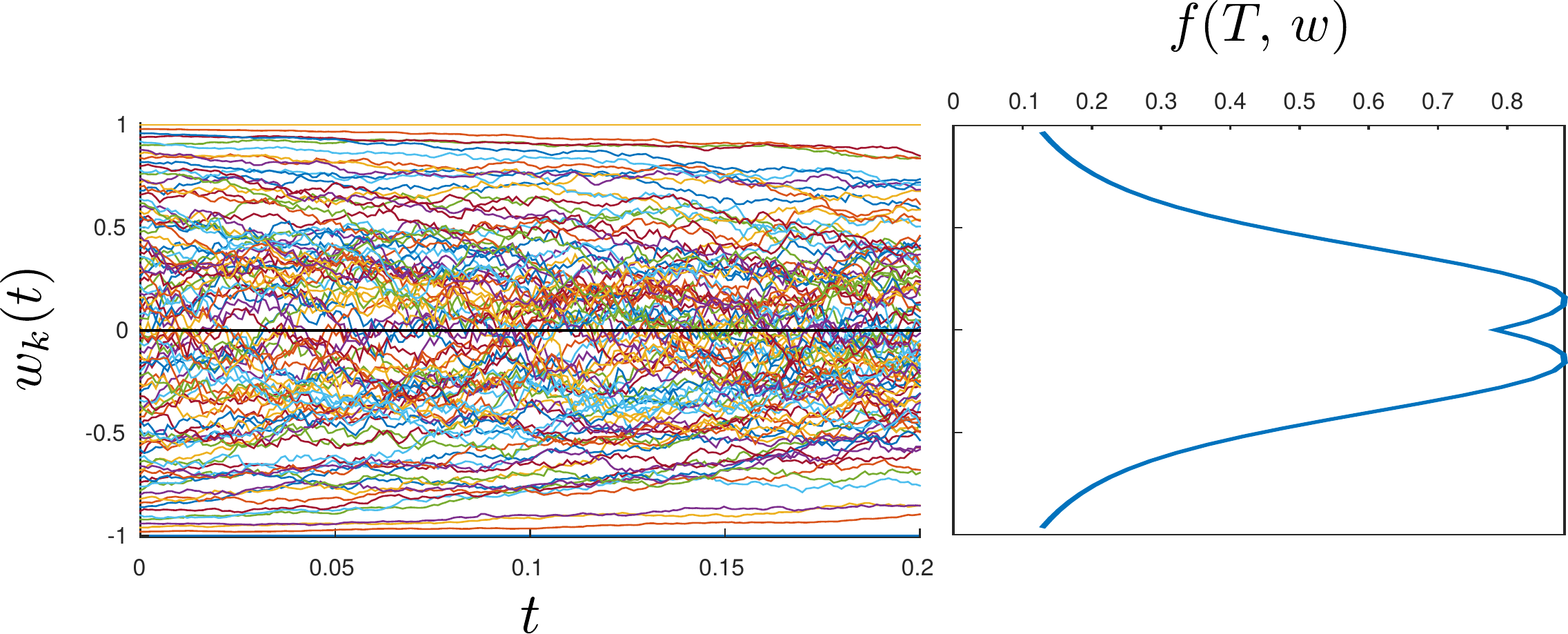}
\caption{Effect of compromise plus self-thinking ($p=0$).}
\label{fig:time_trend-compromise+self-thinking}
\end{figure}

Figure~\ref{fig:time_trend-compromise+self-thinking} shows the result of the simulation when only interactions and self-thinking are included in the model ($p=0$). Here we observe a balance between the trends shown in Figures~\ref{fig:time_trend-only_compromise},~\ref{fig:time_trend-only_self-thinking}.

\begin{figure}[!t]
\centering
\includegraphics[width=\figscale\textwidth]{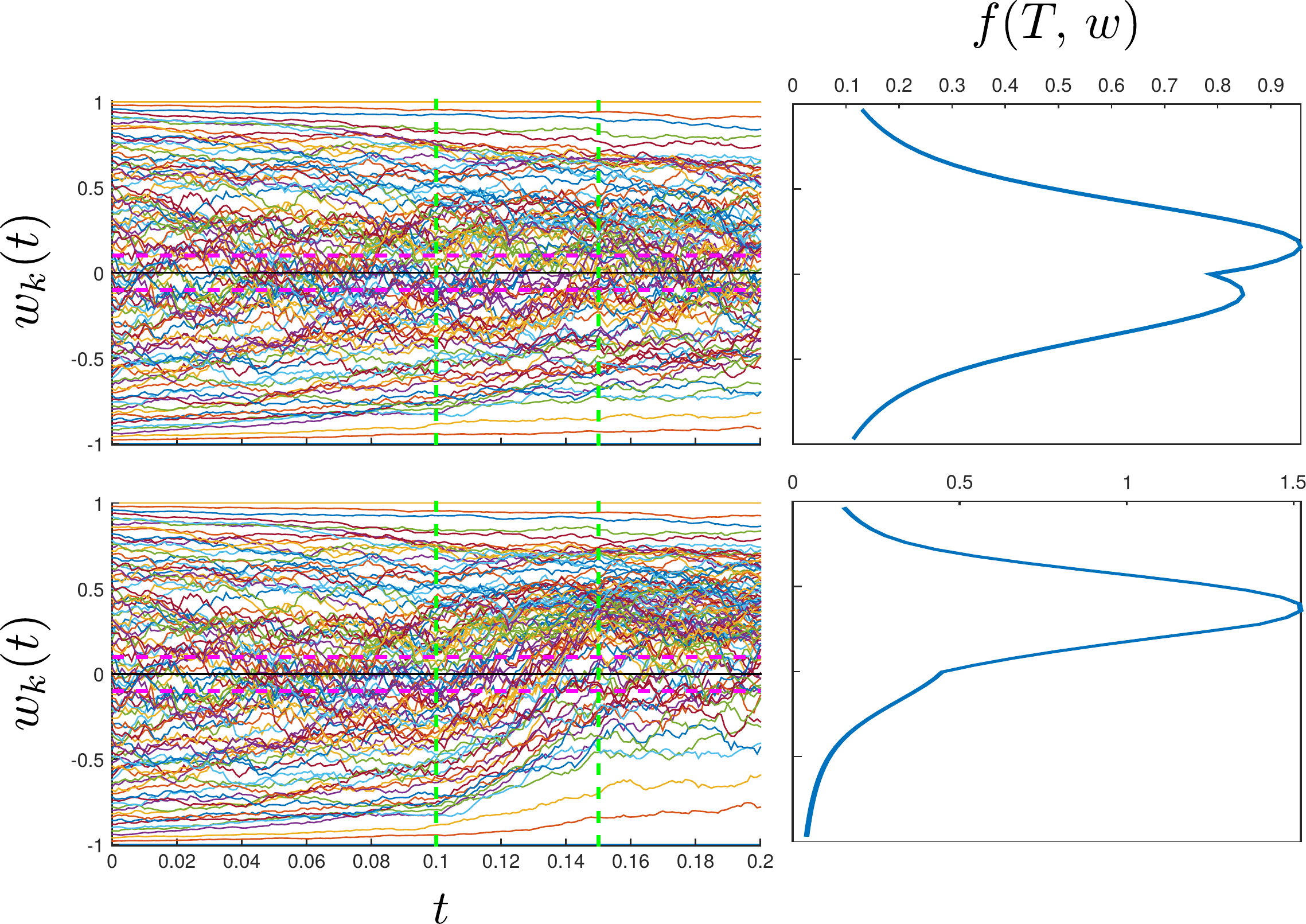}
\caption{Effect of the superposition of compromise, one poll, and self-thinking. Top: the parameters $\alpha$, $p$, $\mu$ are like in Table~\ref{tab:param}. Bottom: the parameter $p$ is increased tenfold. In the left panels, the dashed vertical green lines display the time interval $[T_1,\,T_1+\Delta{T}]$ during which the poll result is influential while the dashed horizontal purple lines identify the opinion interval $[-W^0,\,W^0]$ of the indecisive individuals discarded from the poll result $P_1$.}
\label{fig:time_trend-all}
\end{figure}

Finally, Figure~\ref{fig:time_trend-all} shows the result of the simulation when all the terms are included in the model. We observe here the break of symmetry caused by an opinion poll at time $T_1$ in which the option $\sgn{P_1}=1$ prevails. It is worth pointing out that the continuous model can reproduce the break of symmetry thanks to the coupling with the particle model through $P_1$ in the term $b(t,\,P_1,\,w)$.

\subsection{Scale comparison for increasing $N$}
In the following tests we aim at confirming numerically that the full particle model~\eqref{eq:micro.poll} with $N_P=1$ poll and without hybridization, that is~\eqref{eq:reduced.micro} with $N_\ast=N$ or equivalently~\eqref{eq:hybrid.micro} with $N_\ast=N$ and $\theta=1$, converges to the continuous one~\eqref{eq:fokker-planck} for increasing $N$. To this end we introduce the empirical probability $\Psi_j^m$ of finding the particle opinions in the cell $E_j$, cf.~\eqref{eq:Ej}, at time $t^m\in [0,\,T]$, i.e.,
$$ \Psi_j^m:=\frac{\card\{k\in\{1,\,\dots,\,N\}\,:\,w_k^m\in E_j\}}{N},\qquad j=1,\,\dots,\,J, \quad m=1,\,\dots,\,M, $$
whence we define the empirical probability density of the particle opinions at time $t^m$ as
$$ \Psi^m(w):=\frac{1}{\Delta{w}}\sum_{j=1}^{J}\Psi_j^m\mathbbm{1}_{E_j}(w). $$

\begin{figure}[!t]
\centering
\includegraphics[width=\figscale\textwidth]{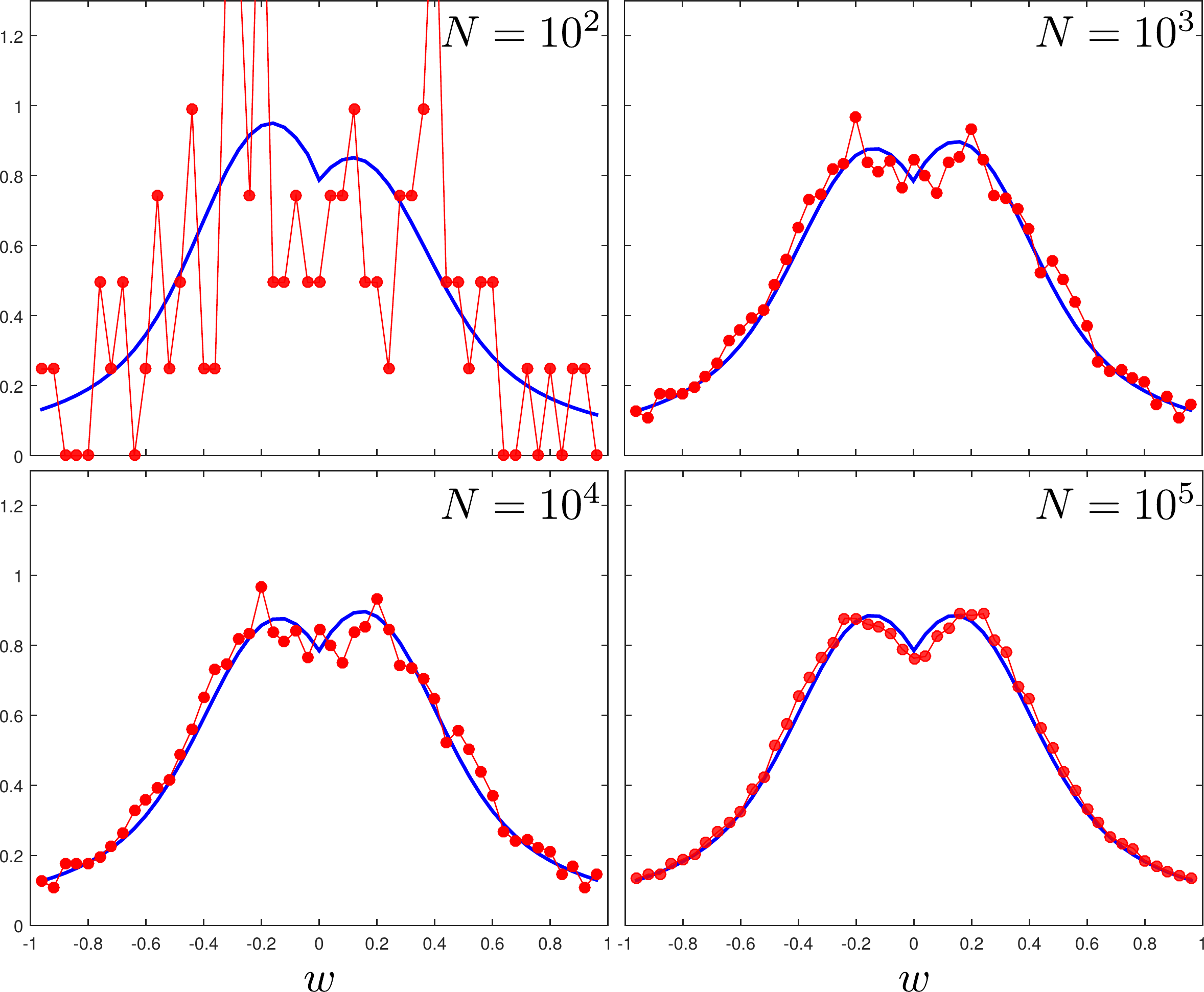}
\caption{Comparison between $f(T,\,w)$ (solid blue line) and $\Psi^M(w)$ (solid red line with bullets) for increasing $N$.}
\label{fig:particle_to_continuous}
\end{figure}

Figure~\ref{fig:particle_to_continuous} compares $f(T,\,w)$ and $\Psi^M(w)$ (recall that $t^M=T$) for increasing orders of magnitude of $N$. It is quite evident that the two slowly converge to one another when $N$ grows.

\begin{figure}[!t]
\centering
\includegraphics[width=\textwidth]{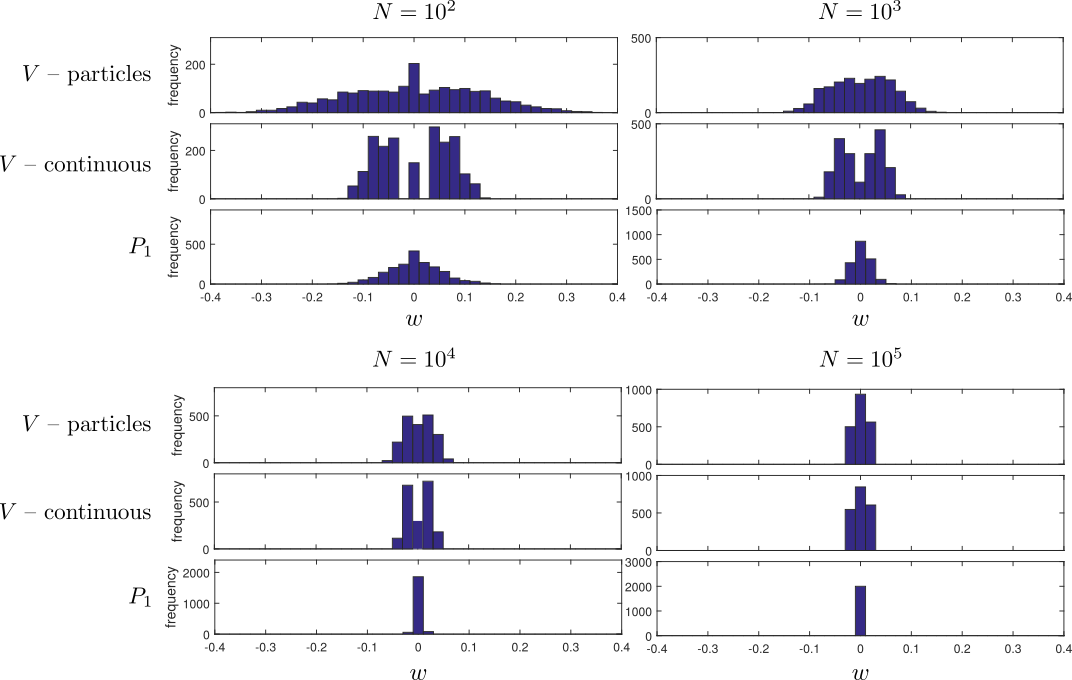}
\caption{Histograms of the final vote $V$ and of the opinion poll $P_1$ for increasing $N$.}
\label{fig:convergence_V}
\end{figure}

As a further test, we also explore the convergence of the statistical distribution of the final vote $V$, cf. Section~\ref{sect:result.vote.micro}. Figure~\ref{fig:convergence_V} shows the histogram of $V$ after $2\cdot 10^3$ runs of the particle model. The analogous quantity is computed also from the continuous model as
$$ V:=\int_0^1 f(T,\,w)\,dw-\int_{-1}^0 f(T,\,w)\,dw, $$
with obviously $V\in[-1,\,1]$. In addition to that, the figure also displays the histogram of $P_1$, i.e., the result of the intermediate poll.

The histograms of $V$ computed from the particle and the continuous model clearly converge to the same profile as $N$ increases. Parallelly, the histogram of $P_1$ tends to a Dirac delta centered in $w=0$, meaning that $P_1^+$ and $P_1^-$ (cf. Section~\ref{sect:opinion.poll}) tend to balance when $N$ is large. Arguably, this is a consequence of the law of large numbers. The convergence of the histograms of $V$ is instead more subtle: since the continuous model depends on the particles through $P_1$, we infer that its trend is comparable with that of the particle model only when $P_1$ approaches its own asymptotic profile, which happens precisely when $N$ is large.

\subsection{Multiscale model}
Now we explore the possibility of reducing the number of agents of the particle model, i.e., to take $N_\ast\ll N$, while still keeping a good agreement with the reference full particle model~\eqref{eq:micro.poll} thanks to the multiscale model discussed in Section~\ref{sect:hybrid.micro.multiscale}. In essence, with the aid of the continuous model we want to recover reliable statistics of the final vote $V$ by tracking a much lower number of individuals than the total size $N$ of the population and using only them to perform both the intermediate polls and the final evaluation of $V$.

In order to compare quantitatively the histograms of $V$ computed with the various models we evaluate the $1$-\emph{Wasserstein distance} of the corresponding normalized profiles. Note that, in this context, the Wasserstein distance is preferable to any $L^p$-distance and that it can be easily computed numerically in one dimension, see e.g.,~\cite{briani2017CMS}.

The reliability of the outcomes of the multiscale model with respect to the reference full particle model strongly depends on the value of the coupling parameter $\theta$. In the following tests we compute numerically, by means of the bisection method, the value of $\theta$ which minimizes the $1$-Wasserstein distance between the histograms of $V$ obtained from the reference full particle model and from the multiscale model, respectively, within a fixed tolerance of $10^{-2}$. In practice we regard the full particle model as an ideal ``exact'' target toward which to lead the more feasible multiscale model, in which only part of the voters are tracked individually.

It is worth stressing that the choice of the best value of $\theta$ remains the main issue of the proposed multiscale technique, since theoretical investigations do not provide so far any \emph{a priori} estimates. Nevertheless, the following tests give some insights into the choice of $\theta$ even if the full particle model cannot be run at all because the number $N$ of particles is prohibitive.

\subsubsection*{Test 1 (One poll, evenly spaced initial opinions)}
\begin{figure}[!t]
\centering
\includegraphics[width=\textwidth]{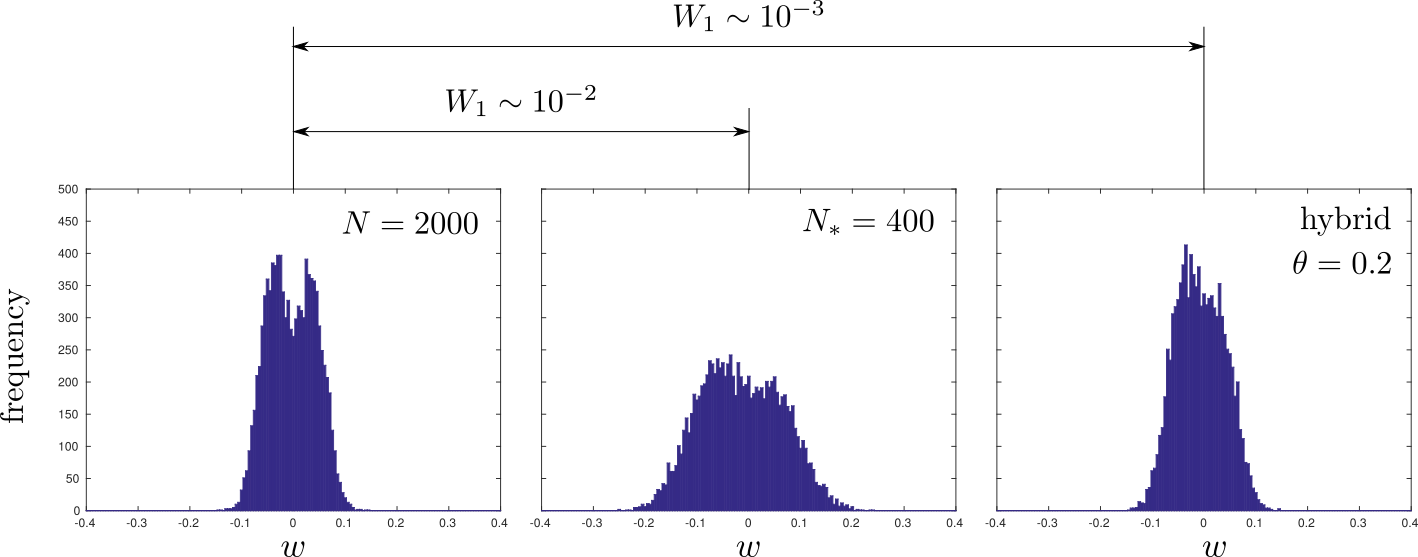}
\caption{Test 1. Histograms of the final vote. Left: full particle model. Center: reduced model. Right: optimally-hybridized particle model ($\theta=0.2$). $W_1$ indicates the $1$-Wasserstein distance between the histograms.}
\label{fig:test1}
\end{figure}

We consider the full particle model with $N=2000$ agents and the reduced model with only $N_\ast=400$ agents. The initial opinions $\{w_k(0)\}$ are taken evenly spaced in $[-1,\,1]$ in both cases. Figure~\ref{fig:test1} shows the histograms of the final vote $V$ after $10^4$ runs for the full particle model, the reduced model alone, and the hybrid particle model with $\theta=0.2$. We stress that in the latter case the histogram of $V$ is still computed out of the $N_\ast$ particle opinions individually tracked by the hybrid particle model, i.e., without considering the contribution of the continuous density $f$. The value $\theta=0.2$ is found by minimizing the Wasserstein distance of the histograms of the reference full particle model and the hybrid particle model. More precisely, the distance between the histograms of the full particle and the reduced model alone is $O(10^{-2})$ while the distance between the histograms of the full particle and the optimally-hybridized particle model is $O(10^{-3})$. Hence the multiscale model allows one to reproduce the target histogram of $V$ more accurately than the simpler reduced model alone, however still tracking only $20\%$ of the total number of particles.

\subsubsection*{Test 2 (One poll, random uniform initial opinions)}
\begin{figure}[!t]
\centering
\includegraphics[width=\textwidth]{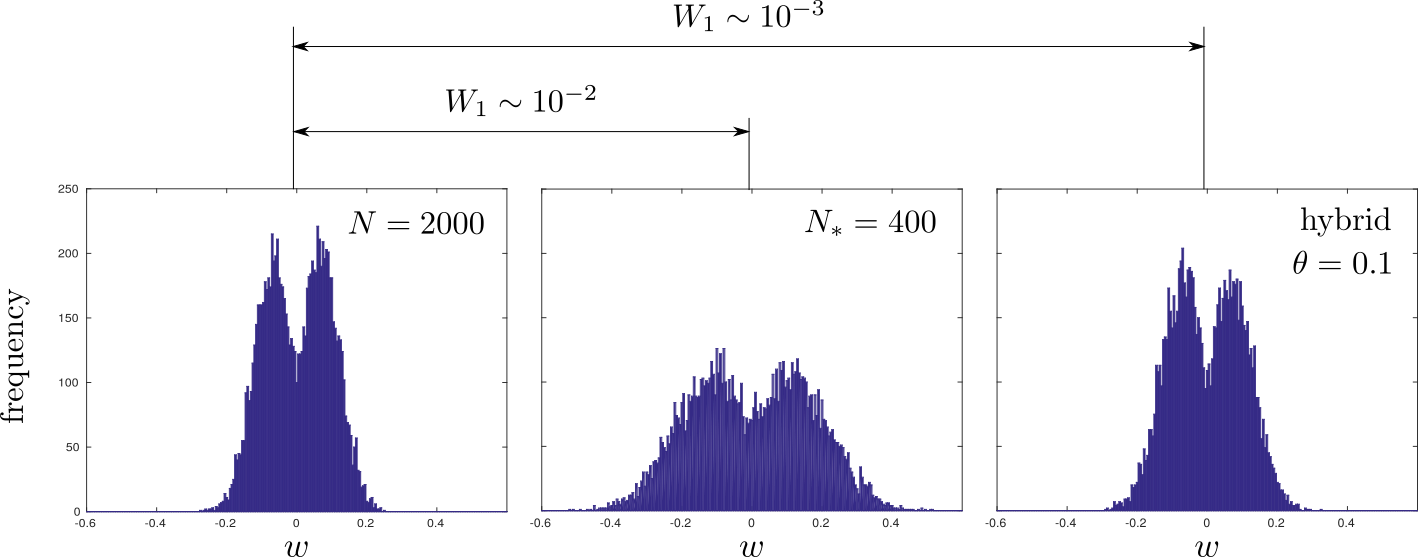}
\caption{Test 2. Histograms of the final vote. Left: full particle model. Center: reduced model. Right: optimally-hybridized particle model ($\theta=0.1$). $W_1$ indicates the $1$-Wasserstein distance between the histograms.}
\label{fig:test2}
\end{figure}

Now we replicate Test 1 with a random uniform distribution of the initial opinions $\{w_k(0)\}\subset [-1,\,1]$ for both the full particle and the reduced model. Figure~\ref{fig:test2} is the counterpart of Figure~\ref{fig:test1} for this case. Notice that the optimal value of $\theta$ which minimizes the Wasserstein distance of the histograms of the reference full particle model and the hybrid particle model is now $\theta=0.1$. In particular, such a distance ($O(10^{-3})$) is again one order of magnitude lower than the distance of the histograms of the full particle model and the reduced model alone ($O(10^{-2})$).

It is worth pointing out that this test differs from the previous Test 1 only in the choice of the $w_k(0)$'s, yet the reference histograms of $V$ are quite different in the two cases (cf. Figures~\ref{fig:test1},~\ref{fig:test2} left). We observe, in particular, that the support of the reference histogram is almost doubled in Test 2, which indicates a wider spread of the possible outcomes of the vote. As a matter of fact no purely continuous model would be able to catch the above difference in the $w_k(0)$'s, for also the evenly spaced distribution in $[-1,\,1]$ of Test 1 is a (very particular) sample from $f_0(w)=\frac{1}{2}\mathbbm{1}_{[-1,\,1]}(w)$. Conversely, by slightly re-tuning $\theta$ the multiscale model succeeds in approximating reliably the reference histogram, thereby confirming that a small amount of particle granularity has to be retained in the dynamics to discriminate between microscopically different scenarios which can give rise to macroscopically observable differences.

\subsubsection*{Test 3 (Two polls)}
\begin{figure}[!t]
\centering
\includegraphics[width=\textwidth]{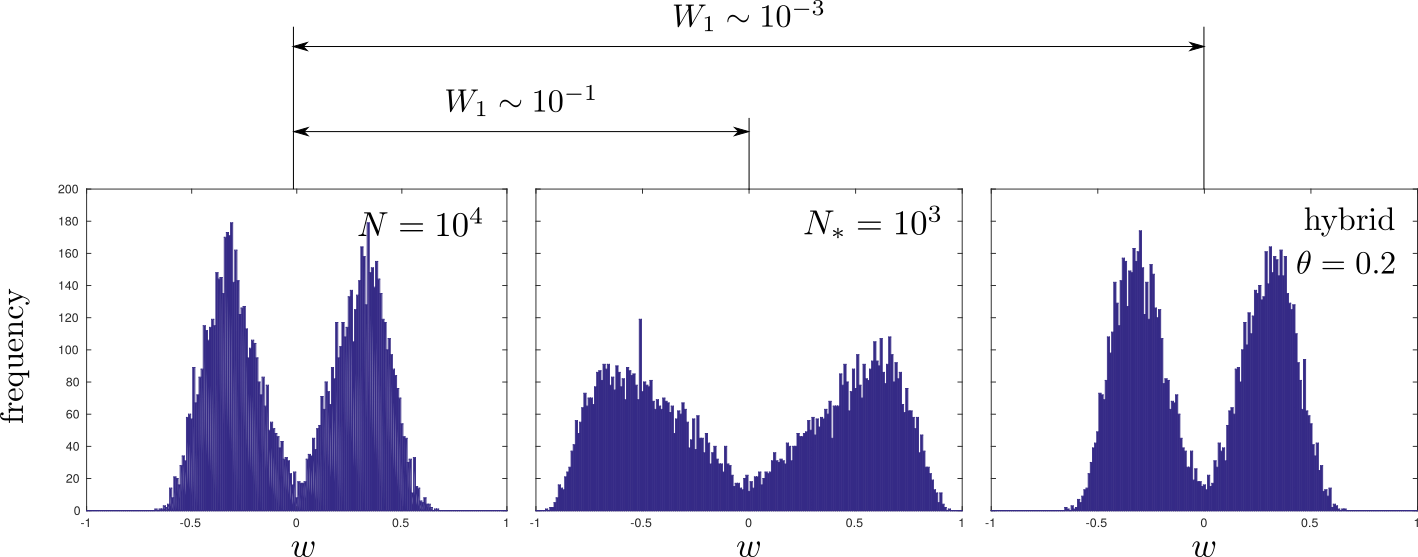}
\caption{Test 3. Histograms of the final vote. Left: full particle model. Center: reduced model. Right: optimally-hybridized particle model ($\theta=0.2$). $W_1$ indicates the $1$-Wasserstein distance between the histograms.}
\label{fig:test3}
\end{figure}

In this test we tackle the more challenging scenario of two polls. We consider a reference full particle model with $N=10^4$ agents and a reduced model with only $N_\ast=10^3$ agents. Figure~\ref{fig:test3} shows the various histograms of the final vote $V$ after $10^4$ runs. As it can be seen, the reduced model alone is considerably far from the reference particle model, while the optimally-hybridized particle model with $\theta=0.2$ gets very close to it. Specifically, the distance of the histograms of the full particle model and the reduced model is $O(10^{-1})$ while the distance between those of the full particle model and the optimally-hybridized particle model is $O(10^{-3})$. Hence in this case the multiscale model allows one to gain two orders of magnitude in the distance from the reference model while still tracking only $10\%$ of the total number of particles.

\begin{figure}[!t]
\centering
\includegraphics[width=\textwidth]{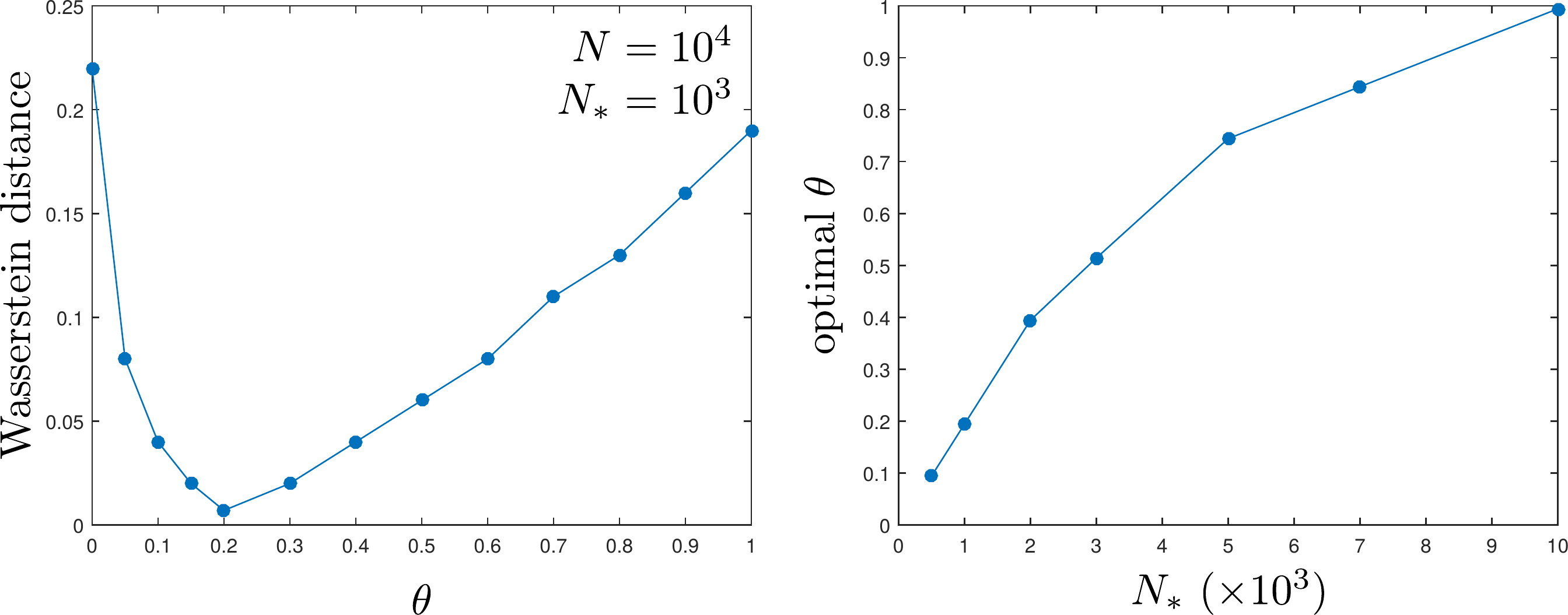}
\caption{Test 3. Left: the $1$-Wasserstein distance between the histograms of $V$ in the full particle model and the hybrid particle model as a function of $\theta$ for fixed $N$, $N_\ast$. Right: the optimal value of $\theta$ as a function of the reduced number $N_\ast$ of agents.}
\label{fig:W1-opt_theta}
\end{figure}

Figure~\ref{fig:W1-opt_theta} shows, on the left-hand side, the Wasserstein distance between the histograms of the full particle model and the hybrid particle model as a function of $\theta$ for this test. As it can be noticed, this function is quite smooth and has in $\theta=0.2$ a unique global minimum. This confirms that the particle granularity cannot be completely neglected in the dynamics. On the right-hand side Figure~\ref{fig:W1-opt_theta} shows instead the optimal value of $\theta$ as a function of $N_\ast$ for this test. As expected, such a function is increasing since the richer of agents the hybrid particle model the smaller the continuous correction it needs. Clearly, if $N_\ast=N$ no correction is needed and the optimal $\theta$ is $1$. Interestingly, from the graph we can also infer that if $N_\ast<500$ then the hybrid particle model cannot be adequately corrected by the Fokker-Planck equation. In fact the distance from the histogram of the full particle model is larger than the selected tolerance $10^{-2}$ for every $\theta\in [0,\,1]$.

\section{Conclusions and future work}
\label{sect:conclusions}
In this paper we have presented a multiscale technique, stemming from the one that we already proposed in~\cite{cristiani2011MMS}, to reduce the number of particles in a multiagent system composed by a usually large but finite number of interacting individuals. The idea is to replace most of the particles with the continuous description of the system formally obtained for an infinite number of agents. A certain much lower number of agents is however still tracked at the particle level in order to preserve the necessary granularity, which can induce small-scale effects of symmetry breaking with potential large-scale impact.

We have applied this technique to a model of opinion dynamics with polls. Here the random and inhomogeneous granular distribution of the opinions is responsible for a break of symmetry which propagates also to the more homogeneous collective dynamics described at the continuous level. Numerical simulations show that, compared to a genuine full particle model, our multiscale approach allows one to reduce by up to $80-90\%$ the number of particles to be tracked explicitly in order to obtain qualitative and quantitative results in line with the expected ones. They also show that the same result can instead not be obtained by either simply reducing the number of particles without multiscale coupling or dropping completely the particles and sticking to a purely continuous model. This confirms that a minimum amount of granularity is essential as a constitutive element of the system.

Possible extensions of the present work can take into account space dynamics, like in~\cite{during2015PRSA}, or the presence of social networks among the agents, like in~\cite{albi2016IFIP,albi2017KRM,moreau2005TAC}, which can possibly shape the opinion paths. In particular, the contacts of each individual, which here we have disregarded in favor of an all-to-all interaction scheme, can possibly change in time on the basis of the evolution of the respective opinions. Even more challenging, our multiscale technique may be used for the control of opinion dynamics~\cite{albi2014PTRSA} possibly targeting only few individuals in the spirit of~\cite{albi2016SIAP,bongini2015PREPRINT,caponigro2015M3AS,fornasier2014PTRSA}.

\section*{Acknowledgments}
The authors want to thank Sauro Succi for his valuable ideas, which contributed to develop this paper.

E.C. is member of GNCS (Gruppo Nazionale per il Calcolo Scientifico) of INdAM (Istituto Nazionale di Alta Matematica), Italy.

A.T. is member of GNFM (Gruppo Nazionale per la Fisica Matematica) of INdAM (Istituto Nazionale di Alta Matematica), Italy.

A.T. acknowledges that this work has been written within the activities of a research program funded by ``Compagnia di San Paolo'' (Turin, Italy).

\appendix

\section{Multiscale coupling of the diffusion terms}
\label{app:coupling.diffusion}
In this section we analyze the multiscale coupling of the diffusion terms in~\eqref{eq:hybrid.micro}. The goal is to justify the stochastic form of that coupling in contrast to the deterministic one used for the other terms of the equation. The forthcoming discussion also complements the numerical results presented in~\cite{cristiani2015JCSMD} for a closely related case study.

Let us consider the stochastic differential equation
\begin{equation}
	dW_t=\sqrt{2\mu}a(W_t)dB_t
	\label{eq:self-thinking}
\end{equation}
describing the microscopic self-thinking process of a single agent as introduced in~\eqref{eq:micro}. From the classical It\^{o}'s calculus it is well known that the Fokker-Planck equation associated with~\eqref{eq:self-thinking} is
\begin{equation}
	\partial_t\phi=\mu\partial_w^2(a^2(w)\phi),
	\label{eq:fokker-planck.phi}
\end{equation}
where $\phi=\phi(t,\,w):\R_+\times\R\to\R_+$ is the probability density function of the random variable $W_t$. Since $a(\pm 1)=0$, integrating~\eqref{eq:fokker-planck.phi} with respect to $w\in [-1,\,1]$ yields $\frac{d}{dt}\int_{-1}^1\phi(t,\,w)\,dw=0$. Thus if $\supp{\phi(0,\,\cdot)}\subseteq [-1,\,1]$ the bounds $-1\leq w\leq 1$ are almost surely never violated at every successive time $t>0$.

For technical reasons which will become apparent in the sequel, we are going to assume
\begin{equation}
	\partial_w\log{\phi}(t,\,\cdot)\in L^\infty(-1,\,1), \quad \forall\,t>0.
	\label{eq:integrability.phi}
\end{equation}
Notice that, since $\partial_w\log{\phi}=\partial_w\phi/\phi$, this assumption is met if e.g. $\phi(t,\,\cdot)\in C^1(-1,\,1)$ is uniformly bounded away from zero. The $C^1$-regularity is quite reasonable, considering that~\eqref{eq:fokker-planck.phi} is a parabolic equation, while the uniform boundedness away from zero depends to some extent on the prescribed initial datum.

The multiscale coupling realized in~\eqref{eq:hybrid.micro} corresponds to interpolating the particle diffusion process~\eqref{eq:self-thinking} with its continuous counterpart~\eqref{eq:fokker-planck.phi}, which we rewrite in the flux form:
$$ \partial_t\phi-\mu\partial_w\left[\left(a^2(w)\partial_w\log{(a^2(w)\phi)}\right)\phi\right]=0, $$
whence we identify the transport velocity $-\mu a^2(w)\partial_w\log{(a^2(w)\phi)}$, cf.~\eqref{eq:D}.

Let us start by examining the case of a deterministic interpolation via the parameter $\theta\in [0,\,1]$. We consider then the new stochastic differential equation
$$ dV_t=a(V_t)\left(\theta\sqrt{2\mu}dB_t-(1-\theta)\mu a(V_t)\partial_v\log{(a^2(v)\phi)}\vert_{v=V_t}dt\right) $$
together with the corresponding Fokker-Planck equation for the probability density function $f=f(t,\,v)$ of the random variable $V_t$:
\begin{equation}
	\partial_tf-(1-\theta)\mu\partial_v\left(a^2(v)\partial_v\log{(a^2(v)\phi)}f\right)
		=\mu\theta^2\partial_v^2(a^2(v)f).
	\label{eq:fokker-planck.det}
\end{equation}
We claim that the probability density function $\phi$ is in general \emph{not} a solution to this equation. Setting $f=\phi$ in~\eqref{eq:fokker-planck.det} we obtain in fact, after standard manipulations,
$$ \partial_t\phi=\mu(\theta^2-\theta+1)\partial_v^2(a^2(v)\phi), $$
which, recalling~\eqref{eq:fokker-planck.phi}, is satisfied only if $\theta=0$ or $\theta=1$. This indicates that the process $V_t$ governed by the multiscale dynamics with deterministic coupling is \emph{not} statistically equivalent to $W_t$.

In order to exhibit a multiscale coupling which preserves the statistical equivalence with the original process $W_t$ in~\eqref{eq:self-thinking} we consider now the stochastic interpolation used in Section~\ref{sect:hybrid.micro.multiscale}, i.e.
$$ dV_t=a(V_t)\left(\Theta\sqrt{2\mu}dB_t-(1-\Theta)\mu a(V_t)\partial_v\log{(a^2(v)\phi)}\vert_{v=V_t}dt\right) $$
with $\Theta\sim\text{Bernoulli}(\theta)$. To deal with the random coefficients $\Theta$, $1-\Theta$ in the derivation of the Fokker-Planck equation corresponding to such particle dynamics we adopt a strategy inspired by that of Sections~\ref{sect:binary.interactions},~\ref{sect:fokker-planck.analysis}. We preliminarily approximate the above equation in a short time $\Delta{t}>0$ as
$$ v^\ast=v+a(v)\left(\Theta\sqrt{2\mu\Delta{t}}Z-(1-\Theta)\mu a(v)\partial_v\log{(a^2(v)\phi)}\Delta{t}\right),
	\quad v\in [-1,\,1], $$
where $Z\sim\mathcal{N}(0,\,1)$, then we consider the associated Boltzmann-type equation for the distribution function $f=f(t,\,v)$, cf.~\eqref{eq:boltzmann.f}:
$$ \frac{d}{dt}\int_{-1}^{1}\varphi(v)f(t,\,v)\,dv=\frac{1}{\Delta{t}}\avefences*{\int_{-1}^{1}\left(\varphi(v^\ast)-\varphi(v)\right)f(t,\,v)\,dv}, $$
where $\avefences{\cdot}$ denotes the expectation with respect to the random variables $\Theta$, $Z$ while $\varphi$ is a sufficiently smooth test function with $\varphi(\pm 1)=\varphi'(\pm 1)=0$. Expanding $\varphi(v^\ast)$ around $v$ for small $\Delta{t}$ while recalling that $\avefences{Z^m}=0$ for all odd $m\in\mathbb{N}$, $\avefences{Z^2}=1$ and that $\avefences{\Theta^m}=\theta$, $\avefences{(1-\Theta)^m}=1-\theta$ for all $m\in\mathbb{N}$ we get
\begin{align}
	\begin{aligned}[t]
		\frac{d}{dt}\int_{-1}^{1}\varphi(v)f(t,\,v)\,dv &= -(1-\theta)\mu\int_{-1}^{1}\varphi'(v)a^2(v)\partial_v\log{(a^2(v)\phi(t,\,v))}f(t,\,v)\,dv \\
		&\phantom{=} +\theta\mu\int_{-1}^{1}\varphi''(v)a^2(v)f(t,\,v)\,dv+R(\Delta{t}),
	\end{aligned}
	\label{eq:Boltzmann.p-f}
\end{align}
the remainder being
\begin{align*}
	R(\Delta{t}) &= \frac{(1-\theta)\mu^2}{2}\Delta{t}\int_{-1}^{1}\varphi''(v)a^4(v)\left(\partial_v\log{(a^2(v)\phi(t,\,v))}\right)^2f(t,\,v)\,dv \\
	&\phantom{=} -\frac{(1-\theta)\mu^3}{6}\Delta{t}^2\int_{-1}^{1}\varphi'''(\bar{v})a^6(v)\left(\partial_v\log{(a^2(v)\phi(t,\,v))}\right)^3f(t,\,v)\,dv,
\end{align*}
where $\min\{v,\,v^\ast\}<\bar{v}<\max\{v,\,v^\ast\}$. We observe that
\begin{align*}
	\abs{a^{2m}\left(\partial_v\log{(a^2\phi)}\right)^m} &= \abs{(a^2)'+a^2\partial_v\log{\phi}}^m \\
	&\leq \left(\abs{(a^2)'}+a^2\abs{\partial_v\log{\phi}}\right)^m \\
	&\leq 2^{m-1}\left(\abs{(a^2)'}^m+a^{2m}\abs{\partial_v\log{\phi}}^m\right)
\end{align*}
for all $m\in\mathbb{N}$, where in the last passage we have used Jensen's inequality. Applying this to $R(\Delta{t})$ with $m=2,\,3$ we get
\begin{align*}
	\abs{R(\Delta{t})} &\leq (1-\theta)\mu^2\Delta{t}\norm{\varphi''}\left(\norm{(a^2)'}^2+\norm{a}^4\norm{\partial_v\log{\phi}}^2\right) \\
	&\phantom{\leq} +\frac{2}{3}(1-\theta)\mu^3\Delta{t}^2\norm{\varphi'''}\left(\norm{(a^2)'}^3+\norm{a}^6\norm{\partial_v\log{\phi}}^3\right),
\end{align*}
whence, considering that $(a^2)'=2aa'$ with $a\in W^{1,\infty}(-1,\,1)$ from~\eqref{eq:a} and taking into account also~\eqref{eq:integrability.phi}, $\abs{R(\Delta{t})}\to 0$ when $\Delta{t}\to 0^+$. Therefore, in the limit $\Delta{t}\to 0^+$ we obtain from~\eqref{eq:Boltzmann.p-f} the Fokker-Planck equation
\begin{equation}
	\partial_tf-(1-\theta)\mu\partial_v\left(a^2(v)\partial_v\log{(a^2(v)\phi)}f\right)=\theta\mu\partial_v^2(a^2(v)f).
	\label{eq:fokker-planck.stoc}
\end{equation}
Thanks to $a(\pm 1)=0$ and to~\eqref{eq:integrability.phi}, also this equation implies $\frac{d}{dt}\int_{-1}^{1}f(t,\,v)\,dv=0$, therefore if $\supp{f(0,\,\cdot)}\subseteq [-1,\,1]$ the bounds $-1\leq v\leq 1$ are almost surely never violated.

Equation~\eqref{eq:fokker-planck.stoc} differs from~\eqref{eq:fokker-planck.det} only for the coefficient $\theta$ in place of $\theta^2$ at the right-hand side. Nevertheless, such a little difference is enough to make the probability density function $\phi$ a solution to this equation \emph{for every} $\theta\in [0,\,1]$. In fact it can be easily checked that letting $f=\phi$ reduces~\eqref{eq:fokker-planck.stoc} to~\eqref{eq:fokker-planck.phi}. This proves that the stochastic multiscale coupling makes the process $V_t$ statistically equivalent to the original process $W_t$ as desired.

\bibliographystyle{amsplain}
\bibliography{CeTa-referendum}

\providecommand{\bysame}{\leavevmode\hbox to3em{\hrulefill}\thinspace}
\providecommand{\MR}{\relax\ifhmode\unskip\space\fi MR }
\providecommand{\MRhref}[2]{%
  \href{http://www.ams.org/mathscinet-getitem?mr=#1}{#2}
}
\providecommand{\href}[2]{#2}
\begin{thebibliography}{10}

\bibitem{albi2016SIAP}
G.~Albi, M.~Bongini, E.~Cristiani, and D.~Kalise, \emph{Invisible control of
  self-organizing agents leaving unknown environments}, SIAM J. Appl. Math.
  \textbf{76} (2016), no.~4, 1683--1710.

\bibitem{albi2015CMS}
G.~Albi, M.~Herty, and L.~Pareschi, \emph{Kinetic description of optimal
  control problems and application to opinion consensus}, Commun. Math. Sci.
  \textbf{13} (2015), no.~6, 1407--1429.

\bibitem{albi2017CHAPTER}
G.~Albi, L.~Pareschi, G.~Toscani, and M.~Zanella, \emph{Recent advances in
  opinion modeling: control and social influence}, Active Particles, Volume 1
  -- Advances in Theory, Models, and Applications (N.~Bellomo, P.~Degond, and
  E.~Tadmor, eds.), Modeling and Simulation in Science, Engineering and
  Technology, Birkh\"{a}user, Basel, 2017.

\bibitem{albi2014PTRSA}
G.~Albi, L.~Pareschi, and M.~Zanella, \emph{Boltzmann-type control of opinion
  consensus through leaders}, Phil. Trans. R. Soc. A \textbf{372} (2014),
  no.~2028, 20140138/1--18.

\bibitem{albi2016IFIP}
\bysame, \emph{On the optimal control of opinion dynamics on evolving
  networks}, System Modeling and Optimization (L.~Bociu, J.-A.
  D\'{e}sid\'{e}ri, and A.~Habbal, eds.), IFIP Advances in Information and
  Communication Technology, vol. 494, Springer, Cham, 2016, pp.~58--67.

\bibitem{albi2017KRM}
\bysame, \emph{Opinion dynamics over complex networks: kinetic modelling and
  numerical methods}, Kinet. Relat. Models \textbf{10} (2017), no.~1, 1--32.

\bibitem{aydogdu2017CHAPTER}
A.~Aydo\u{g}du, M.~Caponigro, S.~McQuade, B.~Piccoli, N.~Pouradier~Duteil,
  F.~Rossi, and E.~Tr\'{e}lat, \emph{Interaction network, state space and
  control in social dynamics}, Active Particles, Volume 1 -- Advances in
  Theory, Models, and Applications (N.~Bellomo, P.~Degond, and E.~Tadmor,
  eds.), Modeling and Simulation in Science, Engineering and Technology,
  Birkh\"{a}user, Basel, 2017.

\bibitem{bennaim2005EL}
E.~Ben-Naim, \emph{Opinion dynamics: rise and fall of political parties},
  Europhys. Lett. \textbf{69} (2005), no.~5, 671--677.

\bibitem{bongini2015PREPRINT}
M.~Bongini, M.~Fornasier, F.~Rossi, and F.~Solombrino, \emph{Mean-field
  {P}ontryagin maximum principle}, Preprint (arXiv:1504.02236), 2015.

\bibitem{briani2017CMS}
M.~Briani, E.~Cristiani, and E.~Iacomini, \emph{Sensitivity analysis of the
  {LWR} model for traffic forecast on large networks using {W}asserstein
  distance}, Preprint (arXiv:1608.00126), 2017.

\bibitem{brugna2015PRE}
C.~Brugna and G.~Toscani, \emph{Kinetic models of opinion formation in the
  presence of personal conviction}, Phys. Rev. E \textbf{92} (2015), no.~5,
  052818/1--9.

\bibitem{cacace2017M2AN}
S.~Cacace, E.~Cristiani, and R.~Ferretti, \emph{Blended numerical schemes for
  the advection equation and conservation laws}, ESAIM Math. Model. Numer.
  Anal. \textbf{51} (2017), no.~3, 997--1019.

\bibitem{caponigro2015M3AS}
M.~Caponigro, M.~Fornasier, B.~Piccoli, and E.~Tr\'{e}lat, \emph{Sparse
  stabilization and control of alignment models}, Math. Models Methods Appl.
  Sci. \textbf{25} (2015), no.~3, 521--564.

\bibitem{carrillo2010SIMA}
J.~A. Carrillo, M.~Fornasier, J.~Rosado, and G.~Toscani, \emph{Asymptotic
  flocking dynamics for the kinetic {C}ucker-{S}male model}, SIAM J. Math.
  Anal. \textbf{42} (2010), no.~1, 218--236.

\bibitem{carrillo2010MSSET}
J.~A. Carrillo, M.~Fornasier, G.~Toscani, and F.~Vecil, \emph{Particle,
  kinetic, and hydrodynamic models of swarming}, Mathematical Modeling of
  Collective Behavior in Socio-Economic and Life Sciences (G.~Naldi,
  L.~Pareschi, and G.~Toscani, eds.), Modeling and Simulation in Science,
  Engineering and Technology, Birkh\"{a}user, Boston, 2010, pp.~297--336.

\bibitem{castellano2009RMP}
C.~Castellano, S.~Fortunato, and V.~Loreto, \emph{Statistical physics of social
  dynamics}, Rev. Mod. Phys. \textbf{81} (2009), 591--646.

\bibitem{ceragioli2016PREPRINT}
F.~Ceragioli and P.~Frasca, \emph{Consensus and disagreement: the role of
  quantized behaviours in opinion dynamics}, Preprint (arXiv:1607.01482).

\bibitem{frasca2012NARWA}
\bysame, \emph{Continuous and discontinuous opinion dynamics with bounded
  confidence}, Nonlinear Analysis: Real World Applications \textbf{13} (2012),
  1239--1251.

\bibitem{chowdhury2016CDC}
N.~R. Chowdhury, I.-C. Mor\u{a}rescu, S.~Martin, and S.~Srikant,
  \emph{Continuous opinions and discrete actions in social networks: A
  multi-agent system approach}, Proceedings of the 55th {IEEE} {C}onference on
  {D}ecision and {C}ontrol, 2016, pp.~1739--1744.

\bibitem{cristiani2015JCSMD}
E.~Cristiani, \emph{Blending {B}rownian motion and heat equation}, J. Coupled
  Syst. Multiscale Dyn. \textbf{3} (2015), no.~4, 351--356.

\bibitem{cristiani2011MMS}
E.~Cristiani, B.~Piccoli, and A.~Tosin, \emph{Multiscale modeling of granular
  flows with application to crowd dynamics}, Multiscale Model. Simul.
  \textbf{9} (2011), no.~1, 155--182.

\bibitem{cristiani2012CDC}
\bysame, \emph{How can macroscopic models reveal self-organization in traffic
  flow?}, Proceedings of the 51st {IEEE} {C}onference on {D}ecision and
  {C}ontrol, 2012, pp.~6989--6994.

\bibitem{cristiani2014BOOK}
\bysame, \emph{Multiscale {M}odeling of {P}edestrian {D}ynamics}, MS\&A:
  Modeling, Simulation and Applications, vol.~12, Springer International
  Publishing, 2014.

\bibitem{cucker2007TAC}
F.~Cucker and S.~Smale, \emph{Emergent behavior in flocks}, IEEE Trans.
  Automat. Control \textbf{52} (2007), no.~5, 852--862.

\bibitem{deffuant2000ACS}
G.~Deffuant, D.~Neau, F.~Amblard, and G.~Weisbuch, \emph{Mixing beliefs among
  interacting agents}, Adv. Complex Syst. \textbf{3} (2000), 87--98.

\bibitem{during2015PRSA}
B.~D\"{u}ring and M.-T. Wolfram, \emph{Opinion dynamics: inhomogeneous
  {B}oltzmann-type equations modelling opinion leadership and political
  segregation}, Proc. R. Soc. A \textbf{471} (2015), no.~2182, 20150345/1--21.

\bibitem{fornasier2014PTRSA}
M.~Fornasier, B.~Piccoli, and F.~Rossi, \emph{Mean-field sparse optimal
  control}, Philos. Trans. R. Soc. A-Math. Phys. Eng. Sci. \textbf{372} (2014),
  no.~2028, 20130400/1--21.

\bibitem{hegselmann2002JASSS}
R.~Hegselmann and U.~Krause, \emph{Opinion dynamics and bounded confidence
  models, analysis, and simulation}, J. Artif. Soc. Soc. Simulat. \textbf{5}
  (2002), no.~3, 1--33.

\bibitem{krause2000CHAPTER}
U.~Krause, \emph{A discrete nonlinear and non-autonomous model of consensus
  formation}, Communications in Difference Equations (S.~Elaydi, G.~Ladas,
  J.~Popenda, and J.~Rakowski, eds.), Proceedings of the Fourth International
  Conference on Difference Equations, CRC Press, 2000, pp.~227--236.

\bibitem{martins2008IJMP-C}
A.~C.~R. Martins, \emph{Continuous opinions and discrete actions in opinion
  dynamics problems}, International Journal of Modern Physics C \textbf{19}
  (2008), no.~4, 617--627.

\bibitem{moreau2005TAC}
L.~Moreau, \emph{Stability of multiagent systems with time-dependent
  communication links}, IEEE Trans. Automat. Control \textbf{50} (2005), no.~2,
  169--182.

\bibitem{pareschi2013BOOK}
L.~Pareschi and G.~Toscani, \emph{Interacting {M}ultiagent {S}ystems: {K}inetic
  equations and {M}onte {C}arlo methods}, Oxford University Press, 2013.

\bibitem{pareschi2017PREPRINT}
L.~Pareschi and M.~Zanella, \emph{Structure preserving schemes for nonlinear
  {F}okker-{P}lanck equations and applications}, Preprint (arXiv:1702.00088),
  2017.

\bibitem{toscani2006CMS}
G.~Toscani, \emph{Kinetic models of opinion formation}, Commun. Math. Sci.
  \textbf{4} (2006), no.~3, 481--496.

\bibitem{wilks1940POQ}
S.~S. Wilks, \emph{Representative sampling and poll reliability}, Public Opin.
  Q. \textbf{4} (1940), no.~2, 261--269.

\end{thebibliography}

\end{document}